\documentclass[pre,aps,superscriptaddress,floatfix,twocolumn,showpacs]{revtex4}  

\usepackage{graphicx}
\usepackage{amsmath}
\font\tenmib=cmmib10
\font\sevenmib=cmmib10 scaled 800

 2

\font\sc=cmcsc10

\font\ottorm=cmr8
\font\msytw=msbm9 scaled\magstep1

\font\indbf=cmbx10 scaled\magstep2

\font\ottorm=cmr8\font\ottoi=cmmi8\font\ottosy=cmsy8%
\font\ottobf=cmbx8\font\ottott=cmtt8%
\font\ottocss=cmcsc8%
\font\ottosl=cmsl8\font\ottoit=cmti8%
\font\sixrm=cmr6\font\sixbf=cmbx6\font\sixi=cmmi6\font\sixsy=cmsy6%
\font\fiverm=cmr5\font\fivesy=cmsy5
\font\fivei=cmmi5
\font\fivebf=cmbx5%

\def\ottopunti{\def\rm{\fam0\ottorm}%
\textfont0=\ottorm\scriptfont0=\sixrm\scriptscriptfont0=\fiverm%
\textfont1=\ottoi\scriptfont1=\sixi\scriptscriptfont1=\fivei%
\textfont2=\ottosy\scriptfont2=\sixsy\scriptscriptfont2=\fivesy%
\textfont3=\tenex\scriptfont3=\tenex\scriptscriptfont3=\tenex%
\textfont4=\ottocss\scriptfont4=\sc\scriptscriptfont4=\sc%
\textfont5=\tenmib\scriptfont5=\sevenmib\scriptscriptfont5=\fivei
\textfont\itfam=\ottoit\def\it{\fam\itfam\ottoit}%
\textfont\slfam=\ottosl\def\sl{\fam\slfam\ottosl}%
\textfont\ttfam=\ottott\def\tt{\fam\ttfam\ottott}%
\textfont\bffam=\ottobf\scriptfont\bffam=\sixbf%
\scriptscriptfont\bffam=\fivebf\def\bf{\fam\bffam\ottobf}%
\setbox\strutbox=\hbox{\vrule height7pt depth2pt width0pt}%
\normalbaselineskip=9pt\let\sc=\sixrm\normalbaselines\rm}
\let\a=\alpha   \let\g=\gamma  \let\d=\delta \let\e=\varepsilon
\let\z=\zeta  \let\h=\eta    \let\k=\kappa \let\l=\lambda
\let\m=\mu        \let\x=\xi     \let\p=\pi    
\let\s=\sigma \let\t=\tau    \let\c=\chi
   
 \let\D=\Delta  \let\L=\Lambda 
\let\P=\Pi

\def\\{\hfill\break} \let\==\equiv

\let\io=\infty 

\let\0=\noindent

\def\ie{{i.e. }}\def\eg{{e.g. }}

\def\tende#1{\,\vtop{\ialign{##\crcr\rightarrowfill\crcr
 \noalign{\kern-1pt\nointerlineskip} \hskip3.pt${\scriptstyle
 #1}$\hskip3.pt\crcr}}\,}
\def\circage{\lower2pt\hbox{$\,\buildrel > \over {\scriptstyle \sim}\,$}}
\def\otto{\,{\kern-1.truept\leftarrow\kern-5.truept\to\kern-1.truept}\,}

\def\T#1{{#1_{\kern-3pt\lower7pt\hbox{$\widetilde{}$}}\kern3pt}}
\def\VVV#1{{\underline #1}_{\kern-3pt
\lower7pt\hbox{$\widetilde{}$}}\kern3pt\,}
\def\W#1{#1_{\kern-3pt\lower7.5pt\hbox{$\widetilde{}$}}\kern2pt\,}

\def\indica{\leaders \hbox to 0.5cm{\hss.\hss}\hfill}
\def\guida{\leaders\hbox to 1em{\hss.\hss}\hfill}

\def\ul{\underline}

\mathchardef\aa   = "050B
\mathchardef\bb   = "050C
\mathchardef\ggg  = "050D
\mathchardef\xxx  = "0518
\mathchardef\zzzzz= "0510
\mathchardef\oo   = "0521
\mathchardef\lll  = "0515
\mathchardef\mm   = "0516
\mathchardef\Dp   = "0540
\mathchardef\H    = "0548
\mathchardef\FFF  = "0546
\mathchardef\ppp  = "0570
\mathchardef\nn   = "0517
\mathchardef\ff   = "0527
\mathchardef\pps  = "0520
\mathchardef\FFF  = "0508
\mathchardef\nnnnn= "056E

\def\to{\rightarrow}

\def\qed{\raise1pt\hbox{\vrule height5pt width5pt depth0pt}}

\def\indic{\hbox{\raise-2pt \hbox{\indbf 1}}}

\def\NNN{\hbox{\msytw N}} 
 \def\ZZZ{\hbox{\msytw Z}}

\def\ul#1{{\underline#1}}

\def\V0{{\bf 0}}

\def\defin{{\buildrel def\over=}}
\def\V#1{{\underline#1}}
\def\Prob{{\rm Prob}}
\newcommand{\beq}{\begin{equation}}
\newcommand{\eeq}{\end{equation}}
\newcommand{\wh}{\widehat}
\newcommand{\wt}{\widetilde}

\begin{document}

\title{
Fluctuation Relation
beyond Linear Response Theory
}
\author{
A.~Giuliani\footnote{e-mail: alessandro.giuliani@roma1.infn.it}
}
\affiliation{
  Dipartimento di Fisica, Universit\`a di Roma
  {\em La Sapienza}, P. A. Moro 2, 00185 Roma, Italy
}
\affiliation{
         INFN, Universit\`a di Roma
         {\em La Sapienza}, P. A. Moro 2, 00185 Roma, Italy
        }
\author{
F.~Zamponi\footnote{e-mail: francesco.zamponi@phys.uniroma1.it}
}
\affiliation{
  Dipartimento di Fisica, Universit\`a di Roma
  {\em La Sapienza}, P. A. Moro 2, 00185 Roma, Italy
}
\affiliation{
         INFM -- CRS Soft, Universit\`a di Roma
         {\em La Sapienza}, P. A. Moro 2, 00185 Roma, Italy
        }
\author{
G.~Gallavotti\footnote{e-mail: giovanni.gallavotti@roma1.infn.it}
}
\affiliation{
  Dipartimento di Fisica, Universit\`a di Roma
  {\em La Sapienza}, P. A. Moro 2, 00185 Roma, Italy
}
\affiliation{
         INFN, Universit\`a di Roma
         {\em La Sapienza}, P. A. Moro 2, 00185 Roma, Italy
        }
\date{\today}
\begin{abstract}
The Fluctuation Relation (FR) is an asymptotic result on the distribution of
certain observables averaged over time intervals $\t$ as $\t\to\io$
and it is a generalization of the fluctuation--dissipation 
theorem to far from equilibrium systems in a steady state which
reduces to the usual Green--Kubo (GK) relation in the limit of small
external non conservative forces. FR is a theorem
for smooth uniformly hyperbolic systems, and it is assumed to be true
in all dissipative ``chaotic enough'' systems in a steady state. 
In this paper we develop a theory of finite time
corrections to FR, needed to compare the asymptotic prediction
of FR with numerical observations, which necessarily involve 
fluctuations 
of observables averaged over finite time intervals $\t$.
We perform
a numerical test of FR in two cases in which non Gaussian fluctuations are 
observable while GK does not apply and we get a non trivial verification of 
FR that is 
{\it independent of} and {\it different from} linear response theory. 
Our results are compatible with the theory
of finite time corrections to FR, while FR would be {\it observably
violated}, well within the precision of our experiments, if such
corrections were neglected. 
\end{abstract}
\maketitle



\section{Introduction}

\noindent
{\it Anosov systems and the fluctuation theorem} - 
The fluctuation theorem concerns fluctuations of phase space
contraction in reversible hyperbolic (Anosov) systems. If time evolution is
described by a differential equation on phase space $M$: $\dot
x=X(x),\, x\in M$,
or by a map $S: x\rightarrow S(x)$ of $M$ one defines the {\it phase space
contraction} as, respectively, $\sigma(x)=-{\rm \,div\,} X(x)$ or
$\sigma(x)=-\log |\det\partial_x S(x)|$. Reversibility means that there is a
metric--preserving map $I$ of $M$ such that $IS=S^{-1}I$ if $S$ is the
time evolution over a certain time $t$ ({\it e.g.} $t=1$).
{\it If the system is Anosov}, that is if $M$ is compact and $S$ is smooth
and uniformly hyperbolic, see \cite{Ga99,GC95a,GC95b,Ga95c,Ru99}, 
the points $x$ will have a well defined
SRB distribution $\mu_{srb}$, \cite{Ru99}, {\it i.e.} almost all points 
w.r.t. the volume measure will evolve
so that {\it all smooth observables} will have a well defined average
equal to the integral over the SRB distribution. Hence, in particular,
the time average of the function $\sigma(x)$ 
will be asymptotically given by the spatial average w.r.t. 
the SRB distribution. In the case of discrete time maps:
\beq
\label{1}
\sigma_+\defin\lim_{\t\to\io} \frac1\t \sum_{j=0}^{\tau-1} \sigma(S^j(x))
=\int_M \sigma\;d\mu_{srb} \defin \langle\; \sigma \;\rangle_{srb}\eeq
If $\sigma_+ > 0$, let:
\beq
\label{2}
p(x)=\frac{1}{\tau \sigma_+} \sum_{j=0}^{\tau-1} \sigma(S^j(x))
\eeq
Analogous definitions are given 
in the continuous time case.
The function $p(x)$ will have average $\langle\; p\;\rangle_{srb} = 1$
and distribution $\pi_\tau(dp)$ such that
\beq
\label{3}
\pi_\tau(\{p\in \Delta\})= e^{\tau \max_{p\in\Delta}\zeta_\infty(p)+ o(\t)}\;,
\eeq
where the correction at the exponent is $o(\t)$ w.r.t. $\t$ as $\t\to\io$.
The following {\it Fluctuation Relation}, discovered 
in a numerical simulation
in \cite{ECM93} and formulated as a theorem for Anosov systems in
\cite{GC95a}, holds:
\beq
\label{4}
\zeta_\infty(p)=\zeta_\infty(-p)+p \sigma_+ \qquad{\rm for\ all}\ |p|<p^*
\eeq
where $\infty> p^*\ge1$ is a suitable (model dependent) constant that, in
general, is {\it different} from the maximum over $\tau$ and $x$ of
$p(x)$; note also that Eq.~\ref{4} is 
(strictly speaking) meaningless in the {\it equilibrium
cases} in which the system is Hamiltonian and reversibility is the
usual velocity sign change: because of the division by $\sigma_+=0$ in 
Eq.~\ref{2}.

\vskip10pt

\noindent
{\it The chaotic hypothesis} - Hyperbolicity is a paradigm 
for disordered systems similar to the
small oscillations paradigm used for ordered motions: it does not
hold exactly in essentially all the physically interesting
systems. The {\it Chaotic Hypothesis} \cite{Ga99,GC95a,GC95b,Ga04,Ru04}
is that nevertheless
one can assume that chaotic motions 
(in the sense of motions with at least one positive Lyapunov exponent) 
exhibit some average properties of truly
hyperbolic motions. This hypothesis is a natural generalization of 
the ergodic hypothesis,
\ie of the assumption that systems of many particles at equilibrium
are well 
described on average by the microcanonical (or by the Gibbs) 
distribution, even if they are not
really (or they are not proven to be) ergodic.
A consequence of the Chaotic Hypothesis is that
(dissipative) deterministic chaotic reversible motions
should have fluctuations of phase space contraction satisfying Eq.~\ref{4}.

One interesting example of such motions is given by a system of 
$N$ interacting particles
in $d$ dimensions subjected to nonconservative forces and kept in a
stationary state by a {\it reversible mechanical thermostat}.
It will be defined by a differential equation $\dot x=X_E(x)$ where 
$x=(\dot{\ul q},\ul q) \in R^{2dN} \equiv M$ ({\it phase space})
and
\beq
\label{INTRO1}
m \ddot{\ul q} = \ul f(\ul q) + \ul g_E (\ul q) - \ul 
\theta_E (\dot{\ul q},\ul q)
\eeq
where $m$ is the mass of the particles, $\ul f(\ul q)$ describes 
the internal (conservative)
forces between the particles and $\ul g_E (\ul q)$ represents 
the nonconservative
``external'' force acting on the system. Finally, 
$\ul \theta_E (\dot{\ul q},\ul q)$ is a
mechanical force that prevents the system from acquiring energy indefinitely: 
this is why we shall
call it a {\it mechanical thermostat}.
Systems belonging to this class are frequently 
used as microscopic models to describe
nonequilibrium stationary states induced by the application of a driving force
(temperature or velocity gradients, electric fields, etc.) on a
fluid system in contact with a thermal bath, \cite{EM90,Ru04}. 
In this context, the phase space contraction
rate $\sigma(x)$ has been identified (setting $k_B=1$) 
with the {\it entropy production rate} \cite{ECM93,Ga99,Ga04,Ru04}, 
the variable $p(x)$ is defined as
\beq
\label{pdix}
p(x) = \frac{1}{\tau \sigma_+} \int_0^\tau dt \ \sigma(S_t x) 
\eeq
(where $x(t) \equiv S_t x$ is the solution of Eq.~\ref{INTRO1} 
with initial datum $x(0)=x$)
and the fluctuation relation has been successfully tested in 
several numerical simulations
\cite{ECM93,BGG97,BCL98,BPV98,GP99,GRS04,ZRA04}.
Having defined the notion of entropy production rate one can 
define a ``duality'' between 
fluxes $\ul J$ and forces $\ul E$ using $\sigma(x)$ as a 
``Lagrangian'' \cite{Ga04}:
\beq
\label{Jdef}
\ul J(\ul E,x) = \frac{\partial \sigma(x)}{\partial \ul E}
\eeq
In the limit $\ul E \rightarrow \ul 0$, {\it i.e.} close to equilibrium, 
the fluctuation relation
leads to Onsager's reciprocity and to Green-Kubo's formulas for 
transport coefficients
\cite{GR97,Ga96}:
\beq
\label{GK}
\mu_{ij} \equiv \lim_{\ul E \rightarrow \ul 0} \frac{\langle J_i 
\rangle_{\ul E}}{E_j} = 
\int_0^\infty dt \ \langle J_i(t) J_j(0) \rangle_{\ul E=\ul 0}
\eeq

\vskip10pt
\noindent
{\it Gaussian distributions} - 
If the distribution $\pi_\tau(p)$ is Gaussian,
$\pi_\tau(p) \propto \exp \left[ -\tau \frac{(p-1)^2}{2 \delta_\tau^2}\right]$,
from the fluctuation relation one can derive an {\it extension} 
of the Green-Kubo relation, i.e. of 
Eq.~\ref{GK}, to finite forces.

Indeed, the fluctuation relation for a Gaussian distribution implies
that the dispersion $\d_\io^2$ of $p$ around its average (equal to $1$) is 
$\delta_\infty^2 = 2/\s_+$ which is, in such case, an extension of a
Green--Kubo formula to non zero fields. One sees this by considering, for
instance, cases in which $\sigma(x)$ is linear in $E$ (as it will be in
the cases that we study numerically below).
Using time-translation invariance one can show that
\beq
\label{deltainfinito}
\delta^2_\infty = \frac{2}{\sigma_+^2}\int_0^\infty dt \
\langle (\sigma(t)-\sigma_+)(\sigma(0)-\sigma_+) \rangle_{E}
\eeq
and from the fluctuation relation $\delta_\infty^2 \sigma_+ = 2$
\beq
\sigma_+ = \int_0^\infty dt \
\langle (\sigma(t)-\sigma_+)(\sigma(0)-\sigma_+) \rangle_{E}
\eeq
Substituting $\sigma(t)=E J(t)$ in the latter expression, one
obtains the relation
\beq
\label{GKestesa}
\frac{\langle J \rangle_E}{E} = \int_0^\infty dt \
[\langle J(t) J(0) \rangle_E - \langle J \rangle_E^2]
\eeq
valid, {\it subject to the Gaussian assumption}, 
also for $E\ne0$.

The leading order in $E$ of the latter relation 
({\it linear response}) is the Green-Kubo formula
for the equilibrium transport coefficient, Eq.~\ref{GK}.

\vskip10pt
\noindent
{\it Numerical verification of the chaotic hypothesis} -
The simplest check of the applicability of the Chaotic Hypothesis
is a check of the fluctuation relation: of course even if the
check has a positive result this will not ``prove'' the hypothesis but
it will at least add confidence to it. A rather stringent test of the
fluctuation relation would be a check which
{\it cannot be reduced to a kind of Green-Kubo relation}; this  requires
at least one of the two following conditions to be satisfied:
\begin{enumerate}
\item the distribution $\pi_\tau(p)$ is distinguishable from a Gaussian, or
\item deviations from the leading order in $E$ in Eq.~\ref{GKestesa}, 
{\it i.e.}, 
deviations from the Green-Kubo relation, are observed.
\end{enumerate}
This is very hard to obtain in numerical simulations of 
Eq.~\ref{INTRO1} for the following
reasons:
\begin{enumerate}
\item to observe deviations from linearity in Eq.~\ref{GKestesa} one has to
apply very large forces $E$, then $\sigma_+$ is very large and 
it becomes very difficult
to observe the negative values of $p(x)$ that are needed to 
compute $\zeta_\infty(-p)$ in Eq.~\ref{4};
\item deviations from Gaussianity in $\pi_\tau(p)$ 
are observed only for values of $p$ that
differ significantly (of the order of $2 \delta_\io$) from $1$ and,
again, it is very difficult
to observe such values of $p$.
\end{enumerate}
Due to the limited computational resources available in the past decade,
all numerical computations that can be found in the literature
on systems described by Eq.~\ref{INTRO1} found that the measured
distribution $\pi_\tau(p)$ could not be 
distinguished from a Gaussian distribution
in the interval of $p$ accessible to the numerical experiment 
\cite{ECM93,BGG97,BCL98,ZRA04}. 

\vskip10pt

The purpose of the present paper is to test the fluctuation relation, 
in a numerical simulation
of a system described by Eq.~\ref{INTRO1}, for large 
applied force when deviations from linearity can be observed, and
the distribution $\pi_\tau(p)$ is appreciably non-Gaussian.
This has become possible
thanks to the fast increase of computational power in the last decade. However,
it is still very difficult to reach values of $\tau$ which
can be confidently regarded as ``close'' to the asymptotic limit 
$\tau \rightarrow \infty$;
thus to interpret our results we develop a theory of the $o(1)$ 
corrections to the function
$\zeta_\infty(p)$ in order to extract the limiting function 
$\zeta_\infty(p)$ from the
numerical data. Taking into account the latter finite time corrections, 
we successfully test the fluctuation relation for 
non--Gaussian distributions and beyond the
linear response theory.

The paper is organized as follows: section II is devoted to the theory 
of finite $\tau$
corrections to the large deviation function 
$\zeta_\infty(p)$ that is needed in the analysis of
the data; in section III we present the model and the details of the numerical
simulation; in section IV we present the details of the data analysis; finally,
in section V and VI we report the result of our simulations.

\section{Finite time corrections to the Fluctuation Relation}
\label{sec:II}

In the present section we 
describe a strategy to study (in principle constructively) 
the $O(1)$ corrections
in the exponent of Eq.~\ref{3}. The theory we propose will
hold {\it assuming that the time evolution is hyperbolic} 
so that it can be applied to
physical systems only if the chaotic hypothesis is accepted. For
simplicity we consider only the case of discrete time evolution via a
map $S$.

\subsection{SRB measure, symbolic dynamics and statistical mechanics}

We study the distribution of $p$ at fixed $\tau$ via its Laplace
transform ({\it characteristic function}) $z_\tau(\lambda)$:
\beq
\label{FTC2.1}
z_\tau(\lambda) = -\frac{1}{\tau} \log 
\langle e^{-\lambda \sum_{j=0}^{\tau-1}\sigma(S^jx)}\rangle_{srb}
\eeq
The main consequence of the hyperbolicity is, 
\cite{Si68,Si77,Ga95c,GC95a,GC95b,Ga96},
that one can find a symbolic
representation of the points of $M$ in terms of sequences
$\underline{\varepsilon}=(\varepsilon_i)_{i=-\infty}^\infty$ of 
finitely many digits $\varepsilon=1,\ldots,k$
subject only to a simple {\it hard core} restriction, namely
$T_{\varepsilon_j,\varepsilon_{j+1}}\equiv 1$ 
if $T$ is a matrix ({\it compatibility matrix})
with entries $0$ or $1$ and such that 
$T^N_{\varepsilon,\varepsilon'}>0$ for some $N>0$
and all $\varepsilon,\varepsilon'$ ({\it mixing} condition). Moreover in such 
a representation the dynamics becomes simply the left
shift, {\it i.e.} if $\underline{\varepsilon}(x)$ 
represents $x$ then $S(x)$ is represented by
the sequence $\underline{\varepsilon}$ shifted to the left by one unit.

The key remarks are
\begin{enumerate}
\item Smooth observables on phase space can be
represented by {\it short range potentials}: in the case of the
observable $\sigma(x)$ this means that there are functions 
$s_X(\underline{\varepsilon}_X)$
defined for all intervals $X=(a,\ldots,a+2n+1)$ and
$\underline{\varepsilon}_X=(\varepsilon_{a},
\ldots,\varepsilon_{a+2n+1})$, {\it translationally invariant}
$s_X=s_{X+b}$ and {\it exponentially decaying} 
on time scale $\kappa^{-1}$ ({\it i.e.} 
$|s_X(\underline{\varepsilon}_X)|< C e^{-\kappa
n}$ for some $C,\kappa>0$), such that 
\beq
\label{FTC2.2}
\sigma(x)= s(\underline{\varepsilon}(x)),\qquad s(\underline{\varepsilon})=
\sum_{X\circ\, 0}
s_X(\underline{\varepsilon}_X)
\eeq
where the sum is over the intervals $X$ centered at the origin
(noted by $X\circ 0$). 
Another important smooth observable is the {\it
expansion rate} $L(x)$ defined as the logarithm of the determinant
of the linearization matrix $\partial S(x)$ ({\it i.e.} the Jacobian
matrix of the map) restricted  
to the unstable manifold: $L(x)=\log \det\partial S(x)_{u}$. This is also
expressible via an exponentially decaying potential $\Phi$:
\beq
\label{expansionrate}
L(x)=\ell(\underline{\varepsilon}(x)),\qquad
\ell(\underline{\varepsilon})=\sum_{X\circ\,0}
\Phi_X(\underline{\varepsilon}_X)
\eeq
\item The SRB distribution, represented as a distribution over
the (compatible) symbolic sequences $\underline{\varepsilon}$, is a
Gibbs state for the short
range potential $\Phi=(\Phi_X(\underline{\varepsilon}_X))$ defined in
Eq.~\ref{expansionrate}, {\it i.e.}
\beq
\label{15}
\langle F \rangle_{srb}=
\lim_{R\rightarrow\infty} 
\frac{\sum_{\underline{\varepsilon}}e^{-\sum_{X\subset
\Lambda_R}\Phi_X(\underline{\varepsilon}_X)} F(\underline{\varepsilon}')}
{\sum_{\underline{\varepsilon}}e^{-\sum_{X\subset
\Lambda_R}\Phi_X(\underline{\varepsilon}_X)}}
\eeq
where $\Lambda_R=(-R,\ldots,R)\subset \ZZZ$, the sums extend over
compatible configurations $\underline{\varepsilon}=
(\varepsilon_{-R},\ldots,\varepsilon_R)$ 
({\it i.e.} with
$T_{\varepsilon_j,\varepsilon_{j+1}}=1$ for $j=-R,\ldots,R-1$), and
$F(\underline{\varepsilon}')$ is an
arbitrary smooth observable defined on phase space regarded as a
function on the symbolic sequences and evaluated at a sequence
$\underline{\varepsilon}'$  which
extends (rather arbitrarily) $\underline{\varepsilon}$ to an infinite 
compatible sequence by continuing $\V\e$ to the right with a sequence
$\V\e_{>}$
and to the left with a sequence
$\V\e_{<}$
into $\V\e'=(\V\e_{<},\V\e,\V\e_{>})$ 
so that $\V\e_{<}$ depends only on the symbol $\e_{-R}$ and
$\V\e_{>}$ depends only on the symbol $\e_{R}$: see
\cite{Si68,Ga02,GBG04}.
\end{enumerate}
The surprising reduction of the problem of studying the SRB
distribution to that of a {\it Gibbs distribution} for a one
dimensional chain with short range interaction (this is the physical
interpretation of Eq.~\ref{15}) generated the possibility of studying
more quantitatively at least some of the problems of nonequilibrium
statistical mechanics outside the domain of ``nonequilibrium
thermodynamics'', \cite{DGM}. 

\subsection{Finite time corrections to the characteristic function}

The characteristic function $z_\tau(\lambda)$
of $p$, see Eq.~\ref{FTC2.1}, can therefore be
computed as 
\beq
\label{4.1}
e^{-\tau z_\tau(\lambda)}=\lim_{R\rightarrow\infty} 
\frac{\sum_{\underline{\varepsilon}}
e^{-\sum_{X\subset \Lambda_R}\Phi_X(\underline{\varepsilon}_X)
-\lambda \,\sum_{X\circ\, [0,\tau-1]}s_X(\underline{\varepsilon}_X) }}
{\sum_{\underline{\varepsilon}}
e^{-\sum_{X\subset\Lambda_R}\Phi_X(\underline{\varepsilon}_X)}}
\eeq
This means that it is the (limit as $R\rightarrow\infty$ of the) ratio between
the partition functions $Z_R(\Phi)$ of a Gibbs distribution in $\Lambda_R$
with potential $\Phi$
(the denominator) 
and the partition function $Z_R(\Phi,\lambda s)$ with the same potential
{\it modified} in the {\it finite} region $[0,\tau-1]\subset\ZZZ$ by
the addition of a
potential $\lambda s_X(\underline{\varepsilon}_X)$.

The one dimensional systems are very well understood and the above is a
well studied problem in statistical mechanics, known as {\it a
finite size corrections} calculation.  For instance it can be attacked
by {\it cluster expansion}, \cite{GBG04}; this is a technique
to deal with the average of the exponential of a 
spin Hamiltonian which is defined in terms of potentials $\phi_X$
exponentially decaying
with rate $\k$, such as those appearing in the numerator 
and in the denominator of Eq.~\ref{4.1}. It allows us to represent
them as:
\beq \label{cluster}\sum_\ul\e e^{-\sum_{X\subset\L_R}\phi_X(\ul\e_X)}=
e^{-\sum_{X\subset\L_R}\wt\phi_X}\;,\eeq
where $\wt\phi_X$ are new {\it
effective} potentials, explicitly computable in terms of suitable averages 
of products of $\phi_X(\ul\e_X)$'s, and which can be proven to be still 
exponentially decaying with the diameter of $X$ with a rate $0<\k'\le\k$. 

In particular, a representation like Eq.~\ref{cluster} 
allows us to rewrite the
partition function in the denominator of Eq.~\ref{4.1} as:
\beq
\label{4.2}
Z_R(\Phi)=\exp \left[ (2R+1) f_\infty(\Phi) - c_\infty(\Phi)+ 
O(e^{-\kappa'R}) \right]
\eeq
and the one in the numerator as
\beq
\begin{split}
\label{19}
&Z_R(\Phi,\lambda s)=\exp \Big[ (2R+1-\tau) f_\infty(\Phi)+
\tau f_\infty(\Phi+\lambda s) \\ 
&- c_\infty(\Phi)- g_\infty(\lambda)+ O(e^{-\kappa'R}+
e^{-\kappa'\tau}) \Big]
\end{split}
\eeq
Therefore
\beq
\begin{split}
\label{20}
z_\tau(\lambda)&=
f_\infty(\Phi)-f_\infty(\Phi+\lambda s)+
\frac{g_\infty(\lambda)}{\tau}+O(e^{-\kappa'\tau}) \\
&\defin z_\infty(\lambda) 
+\frac{g_\infty(\lambda)}{\tau}+O(e^{-\kappa'\tau})
\end{split}
\eeq 

The function $z_\infty(\l)$ is
convex in $\l$ and the functions $g_\infty(\lambda)$ and
$z_\tau(\lambda)$ are analytic in $\lambda$ (a consequence of
the $1$-dimensionality and of the short range nature of the SRB
distribution): namely
$g_\infty(\lambda)=g^{(1)}_\infty \lambda
+\frac12g^{(2)}_\infty\lambda^2+\ldots$ and
$z_\tau(\lambda)=z^{(1)}_\tau\lambda
+\frac12z^{(2)}_\tau\lambda^2+\ldots$ and the coefficients of
their expansion in a power series of $\lambda$ can be expressed in
terms of correlation functions of $\sigma(x)$. For instance, from
Eq.~\ref{FTC2.1} and using the translational invariance of the SRB
measure, \beq\label{zio}
\begin{split}
z^{(1)}_\tau &= \tau^{-1} \langle \sum_{j=0}^{\tau-1}
\sigma(S^jx)\rangle_{srb} = 
\sigma_+ \\
z^{(2)}_\tau &= \tau^{-1} \left[
\langle \sum_{j=0}^{\tau-1}\sigma(S^jx)\rangle_{srb}^2-
 \langle  \sum_{j=0}^{\tau-1} \sigma(S^jx) 
\sum_{k=0}^{\tau-1}\sigma(S^kx) \rangle_{srb}
\right] \\
&=- \sum_{k=-\tau+1}^{\tau-1} \left[1-\frac{|k|}{\tau} \right] 
\langle \sigma(S^kx) \sigma(x) \rangle_{c} 
\end{split}
\eeq
where
$\langle \sigma(S^kx) \sigma(x) \rangle_{c} =
\langle \sigma(S^kx) \sigma(x) \rangle_{srb} - \sigma_+^2$.
Using Eq.~\ref{20}, 
$g_\infty(\lambda)=\lim_{\tau \rightarrow \infty} 
\tau [ z_\tau(\lambda)-
z_\infty(\lambda) ]$, and the analyticity of
$z_\tau(\lambda)$, we have
$g^{(j)}_\infty=\lim_{\tau \rightarrow \infty} 
\tau [ z^{(j)}_\tau - z^{(j)}_\infty ]$.
Since the connected correlation function 
$\langle \sigma(S^kx) \sigma(x) \rangle_{c}$ {\it decays exponentially}
for $k \rightarrow \infty$, we obtain
\beq
\label{22}
\begin{split}
&g^{(1)}_\infty = 0 \\
&g^{(2)}_\infty = \sum_{k=-\infty}^\infty |k| 
\langle \sigma(S^kx) \sigma(x) \rangle_{c}
\end{split}
\eeq

\subsection{Finite time corrections to $\zeta_\infty(p)$}
\label{sec:IIC}

A direct measurement of $z_\tau(\lambda)$ from
the numerical data is difficult. What is really accessible to 
numerical observation are the quantities
$\frac1\tau\log \pi_\tau(\{p\in\Delta\})$ in Eq.~\ref{3} because the
measured values of $p$ are used to build an histogram obtained by
dividing the $p$--axis into sufficiently small
bins $\D$ and counting how many values of
$p$ fall in the various bins. 
Let us choose the size of the bins $\D$ as 
$|\D|=O(\e_\t/\t)$, with 
$\e_\t$ a small parameter which will be eventually chosen $\e_\t=o(1)$,
see Appendix~\ref{app:A} for a discussion of this point.
Let also $p_\D$ be the center of 
the bin $\D$. An application of a local form of central limit theorem,
discussed in Appendix~\ref{app:A}, shows that
the following asymptotic 
representation of $\pi_\tau(\{p\in\Delta\})$ holds:
\beq\label{24}
\pi_\tau(\{p\in\Delta\})=e^{\t\z_\t(p_\D)}\Big(1+o(1)\Big)\eeq
where $\z_\t(p_\D)$ can be interpolated by an analytic 
function of $p$, satisfying the equation 
\beq\label{23}\z_\t(p)=-z_\t(\l_p)+\l_p p\s_+-\frac{1}{2\t}
\log\left[\frac{2\p}{\t}\Big(-\frac{z_\t''(\l_p)}{\s_+^2}\Big)\right]\eeq
and $\l_p$ is the inverse of $p(\l)=z_\t'(\l)/\s_+$. 

Using the previous equations, we now compute the 
lowest order finite time correction to $\z_\io(p)$ around the maximum. 

We rewrite $\z_\t(p)$ as $\z_\t(p)=\z_\io(p)+\frac{\g_\io(p)}{\t}+O(\frac{1}
{\t^2})$. 
By the analyticity of $\z_\t(p)$, we can 
write $\z_\io(p),\g_\io(p)$ around $p=1$ in the form:
$\zeta_\infty(p) = \frac12\zeta_\infty^{(2)} (p-1)^2 + 
\frac1{3!} \zeta_\io^{(3)} (p-1)^3 + 
\ldots$ and 
$\gamma_\infty(p) = \gamma_\infty^{(0)} + \gamma_\infty^{(1)} (p-1) +
\ldots$.

Up to terms of order $(p-1)^2$ and 
higher in the series for $\g_\io(p)$ we can rewrite:
\beq 
\label{25}
\begin{split}
&\z_\t(p)=\\
&=\z_\io(p)+\frac{\g_\io^{(0)}}\t+
\frac{\g_\io^{(1)}}{\t}(p-1)
+O\left(
\frac{(p-1)^2}{\t}\right)+o\left(\frac1\t\right)=\\
&=\z_\io\left(p+
\frac{\g_\io^{(1)}}{\t\z^{(2)}_\io}\right)
+\frac{\g_\io^{(0)}}\t
+O\left(\frac{(p-1)^2}\t\right)+o\left(\frac1\t\right)\;.
\end{split}
\eeq
Thus, the finite time corrections
to $\z_\io(p)$ around its maximum begin with a shift of the maximum 
at 
\beq
\label{26}
p_0 = 1 - \frac{\gamma^{(1)}_\infty}{\tau \zeta^{(2)}_\infty} + 
o\left(\frac1\tau\right)
\eeq
To apply the latter result we need to compute $\gamma^{(1)}_\infty$ in
terms of observable quantities. And, in order to compute $\gamma^{(1)}_\infty$
we apply Eq.~\ref{23}. First of all, we note that $\l_p$ is determined
by the condition
\beq \label{27} p\s_+=z'_\t(\l_p)=\s_++z_\t''(0)\l_p+O(\l_p^2)\eeq
where we used Eq.~\ref{zio} and Eq.~\ref{22}. Then, 
$\l_p=\frac{\s_+(p-1)}{z_\t''(0)}+O\big((p-1)^2\big)$. 
Substituting this result into Eq.~\ref{23} and equating the
terms of order $O(\frac{p-1}{\t})$ at both sides we find:
\beq \label{28}\g_\io^{(1)}=-\frac{1}{2}\frac{z_\io^{(3)}\s_+}{(z_\io^{(2)})^2}\;.
\eeq
The last equation can also be rewritten 
as:
\beq\label{28a}
\gamma_\io^{(1)} = \frac{\zeta_\io^{(3)}}{2 \zeta_\io^{(2)}}
\eeq
This can be proven recalling that $\z_\io^{(2)}$ and 
$\z_\io^{(3)}$
are derivatives of $\z_\io(p)$ in $p=1$, that 
can be obtained by differentiating w.r.t. $\l$ (two or three times, respectively) 
the definition
$\z_\io\big(z_\io'(\l)/\s_+\big)=-z_\io(\l)+\l z_\io'(\l)$ and computing the 
derivatives in $\l=0$ recalling that $z_\io'(0)/\s_+=1$.
Plugging Eq.~\ref{28a} into Eq.~\ref{26} we finally get
\beq \label{wttau}p_0 = 1 - \frac{\z^{(3)}_\infty}{2\tau 
(\zeta^{(2)}_\infty)^2} + 
o\left(\frac1\tau\right)
\eeq
that is the main result of this section.
The key point is that
the moments $\z_\io^{(2)}$ and $\z_\io^{(3)}$ in Eq.~\ref{wttau} are quantities that 
can be measured 
from our empirical data (within an $O(\t^{-1})$ error). 
We then have a verifiable prediction on the expected shift of the maximum
at finite $\t$. Our data agree very well with this prediction, 
see Fig.~\ref{fig_2}
and corresponding discussion in sec.~\ref{sec:IV} below. 

Substituting Eq.~\ref{wttau} in Eq.~\ref{25},
we finally find:
\beq \label{oohh}
\z_\io(p)=
\h_\t(p)+O\Big(\frac{(p-1)^2}{\t}\Big)+o(\t^{-1})\;,\eeq
where $\h_\t(p)$ is defined as
\beq \label{etatau}
\h_\t(p)\defin =-\frac{\g^{(0)}_\io}{\t}+\z_\t\left(p-\frac{\z_\io^{(3)}}{
2\t(\z_\io^{(2)})^2}\right)\;.\eeq
The key point of the above discussion was the validity of 
Eq.~\ref{24}--\ref{23}; see Appendix~\ref{app:A} for their derivation. 

\subsection{Remarks}
\label{sec:IID}

{\it (1)} The shift away from $1$ 
of the maximum of the function $\zeta_\tau(p)$ at
finite $\tau$, expressed by the second term in Eq.~\ref{etatau},
is due to the asymmetry of the distribution $\pi_\tau(p)$
around the average value $p=1$; consequently, it is proportional, at
leading order in $\tau^{-1}$, to $\zeta_\io^{(3)}$ which is 
indeed a measure of the asymmetry of $\zeta_\io(p)$ around $p=1$.
This shift would be absent in the case of a symmetric distribution
({\it e.g.}, a Gaussian) and for this reason it was not observed in 
previous experiments \cite{ECM93,BGG97,BCL98,ZRA04}.

{\it (2)} The error term in the r.h.s. of Eq.~\ref{24} is $o(1)$
w.r.t. $\t$ and it does not affect the computation of 
$\g_\io(p)$. It is then clear that with a calculation similar to that 
we performed, one can get equations for the coefficients $O(\l^k)$ in the 
exponents of Eq.~\ref{24}; in this way 
one can iteratively construct the whole sequence of coefficients 
$\g_\io^{(k)}$ defining the power series expansion of $\g_\io(p)$.

{\it (3)} In models with continuous time evolution the quantity $\sigma_+$ is
not dimensionless but it has dimensions of inverse time: in such cases
one can imagine that one is still studying a map which maps a system
configuration at a time when some prefixed event happens in the system
(typically a ``collision'') into the next one in which a similar event
takes place. If $\tau_0$ is the average time interval between such events
then $\tau_0\sigma_+$ will play the role played by $\sigma_+$ in the 
discrete time case: it will be the adimensional parameter entering 
the estimates of the error terms. 

Note that the coefficients 
$g_\io^{(k)}$ are of order $\s_+^k$, and their
size is necessarily estimated by (the adimensional) entropy production to
the $k$--th power. Then, in the continuous time case, the choice 
of $\t_0$ affects the estimates of the remainders, because it
affects the size of the adimensional parameter $\t_0\s_+$; 
and the size of the mixing time (that is connected with the estimated range of 
decay of the potentials, see \cite{GBG04}). The natural (and physical) 
choice for $\t_0$ is the mixing time. Consistently with this remark,
at the moment of constructing numerically the 
distribution function for the entropy
production rate averaged over a time $\t$, we will always
consider time intervals of the form
$\t=\t_0 n$, $n\ge 1$, see section~\ref{sec:IIIC} below. 

\section{The model}
\label{sec:III}

We consider a system of $N$ classical particles of equal mass $m$ 
in dimension $d$; they are described by their position $q_i$ and momenta 
$p_i = m \dot q_i$,
$(p_i,q_i) \in R^{2d}, \ i=1, \ldots, N$.
The particles are confined in a cubic box of side $L$ with periodic
boundary conditions.
Each particle is subject to a {\it conservative force}, 
$f_i(\ul q) = - \partial_{q_i} V(\ul q)$, and to a
{\it nonconservative force} $E_i$ that does not depend on
the phase space variables. The force $E_i$ is locally conservative
but not globally such due to periodic boundary conditions.
The {\it mechanical thermostat} is a Gaussian thermostat \cite{EM90},
$\theta_i (\ul p,\ul q) = -\alpha(\ul p, \ul q) \, p_i$, and the
function $\alpha(\ul p, \ul q)$ is defined by the condition that the
total kinetic energy 
$K(\ul p) \equiv \frac1{2m} |\ul p|^2 = \frac1{2m} \sum_i p^2_i$ 
should be a constant
({\it isokinetic ensemble}).
The equations of motion are:
\beq
\label{eqofmotion}
\begin{cases}
&\dot{q}_i = \frac{p_i}{m} \ , \\
&\dot{p}_i = f_i(\ul q) + E_i - \alpha(\ul p,\ul q) \ p_i \ ,
\end{cases}
\eeq
From the constraint $\frac{dK}{dt} = 0$ one obtains
\beq
\label{alphadef}
\alpha(\ul p,\ul q)=\frac{\sum_i E_i \ p_i + 
\sum_i f_i(\ul q) \ p_i}{\sum_i p^2_i} \ .
\eeq

\subsection{Entropy production rate}

The total phase space volume contraction rate for this system is given by:
\beq
\begin{split}
&\sigma(\ul p,\ul q)=-\sum_i \left( \frac{\partial \dot{q}_i}{\partial q_i}
+\frac{\partial \dot{p}_i}{\partial p_i} \right) \\
&=  dN\alpha(\ul p,\ul q) + \sum_i \frac{\partial \alpha}{\partial p_i} p_i
= (dN-1) \ \alpha(\ul p,\ul q) \ .
\end{split}
\eeq
Defining the {\it kinetic temperature}, $T \equiv 2 K(\ul p)/(dN-1)$,
\cite{EM90}, the phase space contraction rate can be rewritten
as
\beq\label{36zz}
\sigma(\ul p, \ul q) = 
\frac{\sum_i E_i \ \dot q_i - \dot V}{T} \ .
\eeq
The first term is the power dissipated by 
the external force divided by the kinetic
temperature, and can be identified with the entropy production rate 
\cite{EM90,ECM93,Ga04}. 
The second term is the total derivative w.r.t. time of the potential energy
divided by the temperature: this term does not affect the validity of the 
Fluctuation Relation in the asymptotic limit $\t\to\io$, as total derivatives
give a contribution $O(\tau^{-1})$ in $p(x)$ \cite{Ga04,Ru00}, hence
they do not contribute to $\zeta_\io$; however it has effect on the 
distribution of fluctuations over a finite time $\t$ and 
its influence on the 
numerical computations has been recently discussed in detail \cite{ZRA04}.
The most convenient thing to do, in order to have a finite time
distribution that 
approximates in the best possible way the asymptotic distribution 
of fluctuations,
is to study the distribution of fluctuations for 
the {\it entropy production rate} $\dot s$, where $\dot s$ is identified with
$\s$ {\it minus} the total derivative term $-\dot V/T$ in Eq.~\ref{36zz}:
\beq \label{epr}\dot s(\ul p, \ul q) = 
\frac{\sum_i E_i \ \dot q_i}{T}
\eeq
From now on we will call $\z_\io(p)$ and $\z_\t(p)$ the distributions
for the fluctuations of the entropy production rate $\dot s$ averaged
over infinite or finite time, respectively. These will be the 
objects we will measure and use from now on.

In order to define the {\it current} $J(x,E)$, 
let us rewrite $E_i = E \, u_i$, where $u_i$
is a (constant) unit vector that specifies 
the direction of the force acting on the $i$-th particle.
Then, according to Eq.~\ref{Jdef},
\beq
\label{Jcolor}
J(\ul p, \ul q) = 
\frac{\partial \sigma}{\partial E} = \frac{\sum_i u_i \, \dot q_i}{T}
\eeq
 
\subsection{Discretization of the equations of motion}

To perform the numerical simulation, one has to write the
equations of motion in a discrete form.
One possibility is to use the {\it Verlet algorithm} \cite{AT87};
for Hamiltonian equations of motion ({\it i.e.}, $\ul E=\ul 0$ and $\alpha=0$)
\beq
\label{simpleeqofmotion}
\begin{cases}
&\dot{q}_i = \frac{p_i}{m} \ , \\
&\dot{p}_i = f_i(\ul q) \ ,
\end{cases}
\eeq
the Verlet discretization has the form
\beq
\begin{cases}
&q_i(t+dt) = q_i(t) + \frac{p_i(t)}{m} dt + \frac12 f_i(t) dt^2 \ , \\
&p_i(t+dt) = p_i(t) + \frac12 \big[ f_i(t) + f_i(t+dt) \big] dt \ ,
\end{cases}
\eeq 
where $dt$ is the {\it time step size}. This discretization ensures that
the error is $O(dt^4)$ on the positions $q_i(t)$ in a single time step.
The implementation of this algorithm on a computer is discussed in
detail in \cite{AT87}.

However, this method requires the forces $f_i(t)$ to depend only on the 
positions and not on the velocities: hence, it has to be adapted 
to Eq.s~\ref{eqofmotion}.
This has been done in the following way.
We write the discretized equations as
\beq
\label{eqdiscrete}
\begin{cases}
q_i(t+dt) &= q_i(t) + \frac{p_i(t)}{m} dt \\
&+ \frac12 \big[ f_i(t) + E_i - \alpha(t) p_i(t) \big] dt^2 \ , \\
p_i(t+dt) &= p_i(t) + E_i +\frac12 \big[ f_i(t) + f_i(t+dt) \\ 
&- \alpha(t) p_i(t) - \alpha(t+dt) p_i(t+dt) \big] dt \ ,
\end{cases}
\eeq 
with the same error as in the standard Verlet discretization.
We store in the computer, at time $t$,
the positions $q_i(t)$, the momenta $p_i(t)$,
the forces $f_i(t)$, and the Gaussian multiplier $\alpha(t)$.
Then, we perform the following operations: 
\begin{enumerate}
\item we calculate the new positions $q_i(t+dt)$ using the first equation;
\item using the new positions we calculate the new forces $f_i(t+dt)$ (the
conservative forces depend only on the positions);
\item we calculate the quantity
$\xi_i = p_i(t) +  E_i +\frac12 \big[ f_i(t) + f_i(t+dt) 
- \alpha(t) p_i(t) \big] dt$ and we observe that 
$p_i(t+dt)$ can be expressed in terms of the (known) $\x_i$ and the (unknown)
$\a(t+dt)$ as
\beq
\label{DISCRE1}
p_i(t+dt)=\frac{\xi_i}{1 - \alpha(t+dt) dt/2} \ ;
\eeq

\item substituting Eq.~\ref{DISCRE1} in the definition of $\alpha(t+dt)$, 
Eq.~\ref{alphadef},
we get a self-consistency equation for $\alpha(t+dt)$, whose solution is
\beq
\label{selfconsist}
\begin{split}
&\alpha(t+dt) = \frac{\alpha_0}{1-\alpha_0 dt /2} \ , \\
&\alpha_0 = \frac{ \sum_i E_i \ \xi_i + \sum_i f_i(t+dt) 
\ \xi_i }{\sum_i \xi_i^2} \ ;
\end{split}
\eeq
\item substituting Eq.~\ref{selfconsist} in Eq.~\ref{DISCRE1} 
we calculate $p_i(t+dt)$.
\end{enumerate}
This procedure allows us to calculate the new positions, 
momenta, forces, and $\alpha$,
at time $t+dt$ according to Eq.s~\ref{eqdiscrete} 
{\it without approximations}, 
defining a map $S$ such that $(\ul p(t+dt),\ul q(t+dt))= S(\ul p(t),\ul q(t))$.

Our ({\it discrete}) dynamical system
will be defined by the map $S(\ul p,\ul q)$
and will approximate the differential equations of motion, 
Eq.~\ref{eqofmotion}, with error $O(dt^4)$ 
for the positions and $O(dt^3)$ for the velocities. \\
The map $S$ satisfies the following properties:
\begin{enumerate}
\item it is {\it reversible}, {\it i.e.} it 
exists a map $I(\ul p, \ul q)$ (simply
defined by $I(\ul p, \ul q)=(-\ul p, \ul q)$)
such that $I S = S^{-1} I$;
\item in the {\it Hamiltonian} case ($\ul E=\ul 0$ and $\alpha=0$, 
Eq.s~\ref{simpleeqofmotion}) it is {\it volume preserving}.
\end{enumerate}
The first property ensures that {\it assuming 
the Chaotic Hypothesis} the Fluctuation Relation holds for the map $S$.
The second property ensures that at equilibrium the discretization 
algorithm conserves the phase space volume.

\subsection{Details of the simulation}
\label{sec:IIIC}

In the simulation, we chose the external force of the form 
$E_i = E \, u_i$, where
the unit vectors $u_i$ were parallel to the $x$
direction but with different orientation: half of them were oriented in 
the positive
direction, and half in the negative direction, {\it i.e.} $u_i = (-1)^i \wh x$,
in order to keep the center of mass fixed.
We considered two different systems, selecting interaction potentials
widely used in numerical simulations (for the purpose
of making easier possible future independent checks and rederivations of our
results):
\begin{enumerate}
\item ({\it model I}) the first investigated system is made by $N=8$ 
particles of equal mass $m$ in $d=2$.
The interaction potential is a sum of pair interactions,
$V(\ul q)= \sum_{i < j} v(|q_i - q_j|)$, 
and the pair interaction is represented by a
WCA potential, {\it i.e.} a 
Lennard-Jones potential truncated at the minimum:
\beq
\nonumber
v(r) =
\begin{cases}
& 4 \epsilon \left[ \left(\frac\sigma r \right)^{12} - 
\left(\frac\sigma r \right)^6 \right] + \epsilon \ ,
\hspace{10pt} r \leq \sqrt[6]{2}\sigma \ ; \\
& 0 \ , \hspace{100pt} r > \sqrt[6]{2}\sigma \ .
\end{cases}
\eeq
The reduced density was $\rho=N\sigma^2/L^2=0.95$
(that determines $L$),
the kinetic temperature was fixed to $T=4\epsilon$
and the {\it time step} to $dt=0.001 t_0$, where 
$t_0 = \sqrt{m\sigma^2/\epsilon}$.
In the following, all the quantities will be 
reported in units of $m$, $\epsilon$
and $\sigma$ ({\it LJ units}). 
This system was already studied in the literature, 
see {\it e.g.} \cite{ECM93,SEC98}.
We investigated different values of the external force $E$ ranging from
$E=0$ to $E=25$.
\item ({\it model II})
the second system is a binary mixture of $N$=20 particles (16 of type A and
4 of type B), of equal mass
$m$, in $d=3$, interacting via the same WCA potential of model I; the pair 
potential is
\beq
\nonumber
v_{\alpha\beta}(r) =
\begin{cases}
& 4 \epsilon_{\alpha\beta} \left[ 
\left(\frac{\sigma_{\alpha\beta}} r \right)^{12}
-  \left(\frac{\sigma_{\alpha\beta}} r \right)^{6} \right] 
+ \epsilon_{\alpha\beta} \ ,\\
& \hspace{100pt} r \leq \sqrt[6]{2}\sigma_{\alpha\beta} \ ; \\
& 0 \ , \hspace{88pt} r > \sqrt[6]{2}\sigma_{\alpha\beta} \ ;
\end{cases}
\eeq
$\alpha$ and $\beta$ are indexes that specify the particle species
($\alpha,\beta \in \{A,B\}$).
The parameters entering the potential are the following: 
$\sigma_{AB}=0.8 \sigma_{AA}$;
$\sigma_{BB}=0.88 \sigma_{AA}$;
$\epsilon_{AB}=1.5 \epsilon_{AA}$;
$\epsilon_{BB}=0.5 \epsilon_{AA}$.
Similar potentials have been studied, \cite{KA94,DmSC04},
as models for liquids in the
supercooled regime ({\it i.e.}, below the melting
temperature).
For this system the {\it LJ units} are
$m$, $\epsilon_{AA}$, and $\sigma_{AA}$; the unit of time
is then $t_0 = \sqrt{ m\sigma_{AA}^2/\epsilon_{AA}}$.
The reduced density was $\rho = N \sigma_{AA}^3 / L^3 = 1.2$ 
and the integration step was $dt=0.001 t_0$.
The unit vectors $u_i$ are chosen such that half of the $A$ 
particles and half of the
$B$ particles have positive force in the $x$ direction, 
and the remaining particles 
have negative force in the $x$ direction.
For this system we investigated different values of external 
force $E \in [0,10]$ and temperature $T \in [0.5,3]$.
\end{enumerate}

For each system and for each chosen value of $T$ and $E$, we simulated a very
long trajectory ($\sim 2 \cdot 10^9 dt$) 
starting from a random initial data; we recall that in both
systems we chose $dt=
0.001 t_0$, $t_0$ being the natural unit time introduced in items (1)
and (2) above. After a 
short transient ($\sim 10^3 dt$), still much bigger than the 
decay time $\t_0$ of self-correlations (that 
appears to be $\t_0= 10^2 dt$),
the system reached stationarity, in the sense that the instantaneous
values of observables (\eg potential energy, Lyapunov exponents)
agree with the corresponding asymptotic values within the statistical error
of the asymptotic values themselves.
After this transient we started recording
values $p_i$, $i=1,\ldots,{\cal N}$, 
of the variable $p(x)$ (defined in Eq.~\ref{pdix}),  
integrating the entropy production rate (Eq.~\ref{epr}) 
on adjacent segments of trajectory of length $\tau_0 = 100 dt = 0.1 t_0$.
Note that the length of the time interval over which we averaged the entropy
production rate was chosen as equal to the mixing time, consistently
with the discussion in Remark~(4) of sec.~\ref{sec:IID}.

In conclusion, from each simulation run at fixed $T$ and $E$
we obtain ${\cal N} \sim 10^7$ values $p_i$ of $p(x)$ which are
the starting point of our data analysis.
The value of $\sigma_+$ is estimated by 
averaging the entropy production rate over the whole trajectory.

From a shorter simulation run we measured also the Lyapunov exponents
of the map $S$ using the standard algorithm of Benettin {\it et al.} 
\cite{SEC98,BGS76}.

\subsection{Remarks}

To conclude this section, we note that the WCA potential has a discontinuity in
the second derivative. Thus, one should be concerned with the possibility
that the error of our discretization is not $O(dt^4)$ over the $q_i$'s 
on a single time step, as it should be for potentials $V \in C^4$.
To check that this is not the case (or that 
at least this does not affect our results) we made two independent tests:
\begin{enumerate}
\item we simulated a system similar to model I but with a potential $V \in C^4$
and we obtained qualitatively the same results;
\item we simulated model I using an 
{\it adaptive step size} algorithm \cite{AT87}; 
this kind of algorithms adapt the step size $dt$ during the simulation
in order to keep constant the difference 
between a single step of size $dt$ and two
steps of size $dt/2$. If the precision of our 
discretization changed at the singular
points of the potential, the time step should change 
abruptly during the simulation,
while we observed a practically constant time step during the simulation.
\end{enumerate}
Hence, we have evidence of the fact that the (isolated) singularities 
of the potentials do not produce
relevant effects on our observations;
this is probably due to the fact that the set of singular points of the total 
potential energy $V(\ul q)$ has zero measure w.r.t. the SRB measure.

\section{Data analysis}
\label{sec:IV}

\begin{figure}[t]
\centering
\includegraphics[width=.50\textwidth,angle=0]{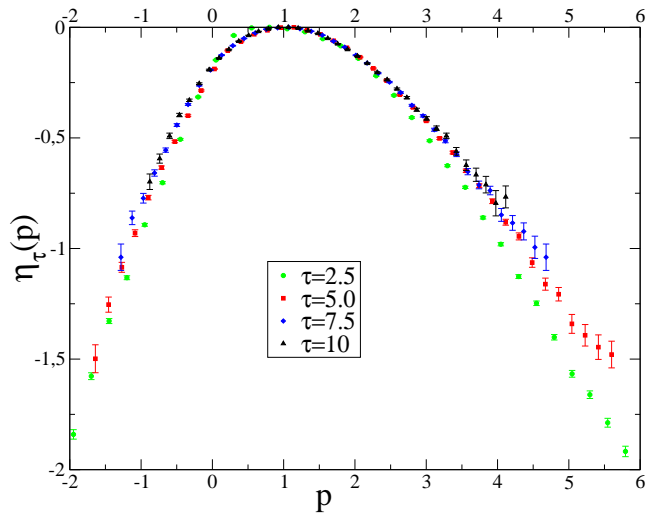}
\caption{Model I at $E=5$: the function 
$\eta_\tau(p)=\zeta_\infty(p)+O((p-1)^2/\tau)$ 
for different
values of $\tau$.}
\label{fig_1}
\end{figure}

In this section we will discuss in detail the procedure we followed to
analyze the numerical data. As an example, we will discuss the 
data obtained from the simulation of model I at $E=5$. As discussed in the
previous section, from the simulation run we obtain a set 
${\cal P}_0 = \{ p_i \}_{i=1\ldots{\cal N}}$ of values of the variable $p(x)$ 
that correspond to $\tau= \tau_0$ and are 
measured on adjacent segments of trajectory.
We recall again that $\tau_0 = 0.1 = 100 dt$ is of the order 
of the {\it mixing time}, {\it i.e.} 
the time scale over which the correlation functions 
({\it e.g.} of density fluctuations) decay to zero.

\vskip10pt

\noindent
{\it Probability distribution function} -
From the dataset ${\cal P}_0$ we construct the histograms 
$\pi_\tau(p)$ for different values of $\tau = n \tau_0$ as follows: 
the values of $p(x)$ for $\tau = n\tau_0$ are obtained by averaging 
$n$ subsequent entries of the dataset ${\cal P}_0$; we obtain a new dataset
${\cal P}_n = \{ p^{(n)}_j \}_{j=1\ldots{\cal N}/n}$ such that 
$p^{(n)}_j= n^{-1} \sum_{i=nj+1}^{n(j+1)} p_i$.
Finally, from the dataset ${\cal P}_n$ the histogram of $\pi_{\tau}(p)$ is
constructed for $\tau=n\tau_0$; 
the errors are estimated as the square roots of the number of counts
in each bin.
The function $\zeta_{\tau}(p)$ is then defined as
$\zeta_\t(p)=\t^{-1} \log \pi_\t(p)$.

\begin{figure}[t]
\includegraphics[width=.50\textwidth,angle=0]{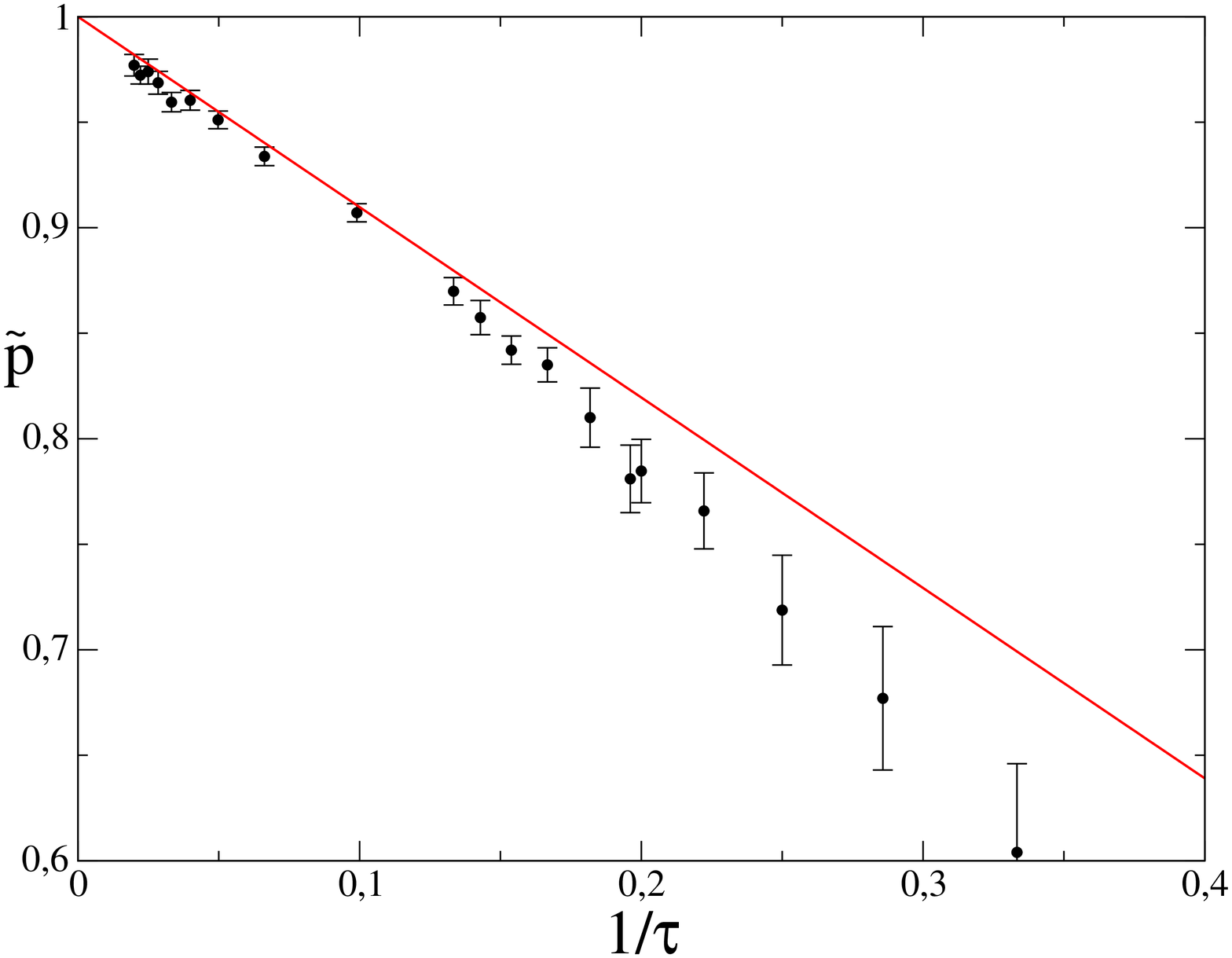}
\caption{Model I at $E=5$: the maximum $\widetilde p_\tau$ of $\zeta_\tau(p)$ 
as a function of $1/\tau$.
The full line is the prediction of Eq.~\ref{wttau}, 
$\widetilde p = 1 - \zeta_\io^{(3)}/\big[2\tau 
\big(\zeta_\io^{(2)}\big)^2\big]$.}
\label{fig_2}
\end{figure}

\vskip10pt

\noindent
{\it Shifting of the maximum} -
By fitting the function $\zeta_{\tau}(p)$ in $p\in[-1,3]$ 
with a sixth-order polynomial we determine the position of the maximum
$\widetilde p_\tau$ within an error that, since $\d p$ is the length of a bin,
we estimate to be $\d p/2$.
Then, we construct the function 
$\eta_{\tau}(p) = \zeta_\tau(p-1+\widetilde p_\tau)$ 
(see Eq.~\ref{etatau}) which
is expected to approximate the limiting function $\zeta_\io(p)$ with error
$O((p-1)^2/\tau)$.
The functions $\eta_\tau(p)$ are reported 
in Fig.~\ref{fig_1} for different
values of $\tau$. 
We observe a very good convergence for $\tau \gtrsim 5.0 = 50\tau_0$.

By a fourth-order fit of the so-obtained limiting function $\zeta_\io(p)$ 
around $p=1$ we extract the coefficients $\zeta_\io^{(2)}=-0.287$ and 
$\zeta_\io^{(3)}=0.149$ in order to test the correctness of 
Eq.~\ref{wttau}.
In Fig.~\ref{fig_2} we report $\widetilde p_\tau$. 
The full line is the prediction of 
Eq.~\ref{wttau}, that is indeed verified for $\tau \gtrsim 10$.
This result confirms the analysis of section~\ref{sec:II}.

\begin{figure}[t]
\includegraphics[width=.50\textwidth,angle=0]{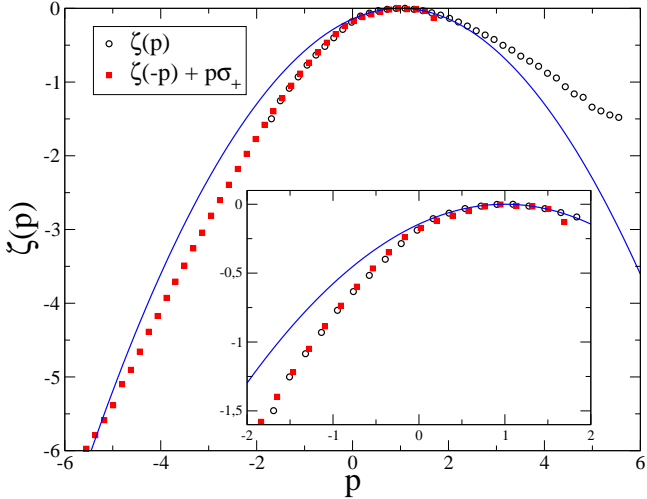}
\caption{Model I at $E=5$: 
the estimate of the function $\zeta_\io(p)$ (open circles). 
In the same
plot $\zeta_\io(-p) + p\sigma_+$ (filled squares) 
is reported. In the inset, the
interval $p\in[-2,2]$ where the data overlap 
is magnified. The full
line is the Gaussian approximation, 
$\frac12\zeta_\io^{(2)}(p-1)^2$.
The plot shows that the Gaussian is not a good approximation
in the interval $[-2,2]$. The validity of the Fluctuation Relation
in the same interval is shown by the overlap of the open circles and 
filled squares.}
\label{fig_3}
\end{figure}

\vskip10pt

\noindent
{\it Graphical verification of the fluctuation relation} - 
From the previous analysis we can conclude that the function $\eta_\tau(p)$ 
for $\tau = 5.0$ provides a good estimate 
of the function $\zeta_\io(p)$ for $p\in[-2,4]$
(see Fig.~\ref{fig_1});
thus, we can use this function to test the 
fluctuation relation, Eq.~\ref{4},
in this range of $p$. In Fig.~\ref{fig_3} we report the estimated functions
$\zeta_\io(p)$ and $\zeta_\io(-p)+p\sigma_+$. An excellent agreement between 
the two functions is observed in the interval $p\in[-2,2]$ where our data 
allows the computation of both $\zeta_\io(p)$ and $\zeta_\io(-p)$.
Note that in this range of $p$ the function $\zeta_\io(p)$ is not Gaussian, see
the inset of Fig.~\ref{fig_3}.

\vskip10pt
\noindent
{\it Quantitative verification of the fluctuation relation} - 
The translation of the function $\zeta_\tau(p)$ is crucial to obtain a
correct estimate of the limit $\zeta_\infty(p)$ and to verify the fluctuation
relation. 
In this section we will try to quantify this observation; 
as the discussion will
be very technical, the reader who is satisfied with Fig.~\ref{fig_3} should 
skip to next section.

\begin{figure}[t]
\includegraphics[width=.33\textwidth,angle=0]{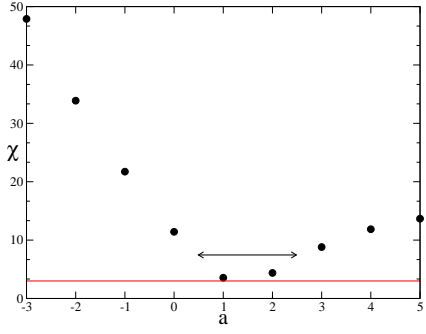}
\caption{Model I at $E=5$: 
the function $\chi(a)$. The full line corresponds to $\chi=3$.
The arrow indicates the interval $\d_0\pm\d p/2$
(note that its length is 2 in units of $a$)
into which the minimum of $\chi$ can be located 
within the accuracy of the histogram.}
\label{fig_4}
\end{figure}

The histogram $\pi_{n\tau_0}(p)$ derived from the 
dataset ${\cal P}_n$ is constructed
assigning the number of counts $\pi_\alpha$ in the 
$\alpha$-th bin to the middle
of the binning interval, that we call $p_\alpha$ 
(the latter will be an {\it increasing} function of $\alpha$).
The statistical error 
$\delta \pi_\alpha$ on the number of counts is $\sqrt{\pi_\alpha}$. 
Our histograms are constructed
in such a way that if $p_\alpha$ is the center 
of a bin, also $-p_\alpha$ is the center
of a bin; we call $\overline\alpha$ the bin 
such that $p_{\overline\alpha}=-p_\alpha$.
There exists a value $p_m$ such that for $p_\alpha < p_m$ the
number of counts in the bin $\alpha$ is smaller than $m$ (we choose $m=4$).
Let us indicate by $p_{\alpha_m}$ the smallest value of $p_\alpha > p_m$. 
Hence, the histogram is characterized by:
\begin{enumerate}
\item a {\it bin size} $\delta p$;
\item
the bin $\alpha_m$ corresponding to the minimum value of $p_\alpha$ such 
that the number of counts in the bin is at least $m$;
\item the total number $M$ of bins such that 
$\alpha \in [\alpha_m,\overline\alpha_m]$; for these values of $p_\alpha$,
both $\pi_\tau(p)$ and $\pi_\tau(-p)$ 
can be computed and they can be used to verify
the fluctuation relation.
\end{enumerate}
The function $\zeta_{\tau}(p)$, derived from the histogram, is specified by
a set of values $(p_\alpha,\zeta_\alpha,\delta \zeta_\alpha)$ 
for each bin $\alpha$,
where $\zeta_\alpha=\tau^{-1} \log \pi_\alpha$ and 
the error $\delta\zeta_\alpha$ has been defined by
\beq
\delta \zeta_\alpha = \frac1\tau \frac{\delta \pi_\alpha}{\pi_\alpha}=
\frac1{\tau \sqrt{\pi_\alpha}} \ .
\eeq
A quantitative verification of Eq.~\ref{4} 
is possible defining the following
$\chi^2$ function:
\beq
\chi^2 \equiv \frac1M \sum_{\alpha=\alpha_m}^{\overline\alpha_m} 
\frac{(\zeta_\alpha - \zeta_{\overline\alpha} - p_\alpha \sigma_+)^2}
{(\delta \zeta_\alpha)^2 + (\delta \zeta_{\overline\alpha})^2}
\eeq
The value of $\chi$ is the average difference between $\zeta_\tau(p)$ and
$\zeta_\tau(-p)+p\sigma_+$ in units of the statistical error.
Translating $p$ of a quantity $a \delta p/2$, $a\in \ZZZ$,
corresponds to shifting the histogram, {\it i.e.} to
consider a new histogram $(p_\alpha + a \delta p/2,\zeta_\alpha,
\delta\zeta_\alpha)$.
This preserves the property that if $p_\alpha$ is the center of a bin, also
$-p_\alpha$ is the center of a bin; we
call $\overline\alpha(a)$ the new value of $\alpha$
such that $p_{\overline\alpha(a)}+a \delta p/2=
-(p_\alpha + a \delta p/2)$. Also, the number $M_a$ of bins such that
$\alpha(a) \in [\alpha_m, \overline\alpha_m(a)]$ depends on $a$.
We define
\beq
\label{chia}
\chi^2(a) \equiv \frac1{M_a} \sum_{\alpha=\alpha_m}^{\overline\alpha_m(a)} 
\frac{\big(\zeta_\alpha - \zeta_{\overline\alpha(a)} - 
(p_\alpha + a\delta p/2) \sigma_+\big)^2}
{(\delta \zeta_\alpha)^2 + (\delta \zeta_{\overline\alpha(a)})^2}
\eeq
We shall use the criterion that 
the fluctuation relation is satisfied if $\chi \leq 3$, which means that
$\zeta_\io(p)$ and $\zeta_\io(-p)+p \sigma_+$
differ, {\it on average}, by less than $3$ times the statistical error 
$\sqrt{ \big( \delta \zeta(p)\big)^2 +\big( \delta \zeta(-p)\big)^2}$.
The function $\chi(a)$ for the case of model I at $E=5$ 
is reported in Fig.~\ref{fig_4}. 

The minimum of $\chi$
is assumed between $a^*=1$ and $a^*+1=2$ and an upper limit for the value of 
$\chi$ at the minimum is $\chi(1)=3.5$.
We estimate the translation that minimizes $\chi$ as 
$\delta_0 = (a^*+0.5) \delta p/2 = 1.5 \cdot 0.093 = 0.140$, 
and to this estimate
we attribute an error $\pm\d p/2$, where $\delta p = 
0.186$ is the size of a bin. 
On the other hand, we have seen above that,
in order to shift the maximum of $\zeta_\tau(p)$ 
in $p=1$, one has to translate $p$ by a 
quantity $\delta \equiv 1 - \widetilde p = 0.215$. 
The consistency of our analysis
requires that $\delta$ and $\d_0$ coincide within their errors, \ie 
that the intervals $\d\pm\d p/2$ and $\d_0\pm\d p/2$ overlap, or in other
words $|\d-\d_0|<\d p$. In the present case $0.075=|\d-\d_0|<\d p=0.186$,
then $\d$ and $\d_0$ coincide within the errors. This means that
the translation of $p$ 
brings the maximum of $\zeta_\tau(p)$ in
$p=1$ and, {\it at the same time}, minimizes the difference between
$\h_\tau(p)$ and $\h_\tau(-p)+p \sigma_+$, where $\h_\t$ is our
finite time estimate of $\z_\io(p)$. 
The value $\chi(a^*)$ quantifies
this difference and is a first estimate of the precision of our analysis.

\begin{figure}[t]
\includegraphics[width=.33\textwidth,angle=0]{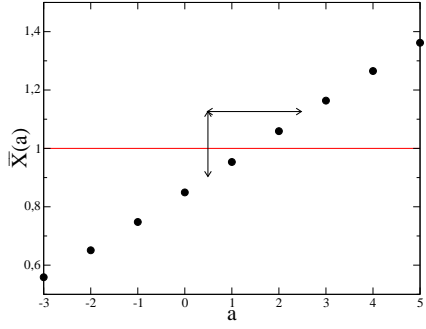}
\caption{Model I at $E=5$: the function $\overline X(a)$. 
The horizontal arrow marks the interval where 
the minimum of $\chi$ is located, see
Fig.~\ref{fig_4}.
The vertical arrow indicates the error 
$\delta X$ on the value $X=1$ which is estimated
as $\delta X=2(\overline X(2)-\overline X(1))$. 
The optimal slope of the fluctuation relation without
the translation would have been $\overline X(a=0) \sim 0.85$.}
\label{fig_5}
\end{figure}

Another estimate of the precision of our analysis can be obtained as follows. 
We define a parameter $X$ as the slope of $\zeta_\io(p)-\zeta_\io(-p)$ as a
function of $p\sigma_+$:
\beq
\zeta_\io(p)= \zeta_\io(-p)+Xp\sigma_+
\eeq
The fluctuation theorem predicts $X=1$, but other values of $X$ are possible 
under different hypothesis, see \cite{BG97,BGG97,Ga04,Ga99b}.
We define a function $\chi^2(a,X)$ as
\beq
\label{chiaX}
\chi^2(a,X) \equiv \frac1{M_a} 
\sum_{\alpha=\alpha_m}^{\overline\alpha_m(a)} 
\frac{\big(\zeta_\alpha - \zeta_{\overline\alpha(a)} - 
X (p_\alpha+a \delta p /2) \sigma_+\big)^2}
{(\delta \zeta_\alpha)^2 + (\delta \zeta_{\overline\alpha(a)})^2}
\eeq
and for each value of $a$ we calculate the optimal value of 
$X$, $\overline X(a)$,
by minimizing $\chi^2(a,X)$. The function $\overline X(a)$ is reported in 
Fig.~\ref{fig_5}. As the shift of the maximum $\delta$ is between $a=1$ and
$a=2$, we see that the slope $X$ is compatible with one. Moreover, as the
natural error on $p$ is the size of a bin $\delta p$, we assign to the
value $X=1$ a statistical error 
$\delta X= 2 ( \overline X(2)-\overline X(1) ) = 0.22$.
Note again that without the translation of $p$ the optimal slope would be
$X \sim 0.85$, incompatible with Eq.~\ref{4}.

\vskip10pt
\noindent
{\it Discussion} - From the present analysis, we can conclude that:
\begin{enumerate}
\item the translation shifting the maximum of 
$\z_\t(p)$ to $p=1$ at the same time
minimizes the difference between $\h_\tau(p)$ and $\h_\tau(-p)+p \sigma_+$, 
where $\h_\t$ is our
finite time estimate of $\z_\io$; this proves the consistency of our
theory of finite time corrections;
\item without the translation of $p$ (that corresponds to $a=0$), 
the function $\zeta_\tau(p)$ for $\tau \sim 5.0$
{\it do not satisfy the fluctuation relation}, as $\chi(a=0) = 11$ 
and $\overline X(a=0)=0.85$;
\item the function $\eta_\tau(p)=\zeta_\tau(p-\delta)$ 
satisfies the fluctuation relation
with $\chi \sim 3$ and an error of about $20\%$ on the slope $X$: 
both quantities measure
the accuracy of our analysis.
\end{enumerate}
Thus, the check of the fluctuation relation relies 
crucially on the translation
of the function $\zeta_\tau(p)$ that has been discussed 
in section~\ref{sec:II}.
By considering larger values of $\tau$ one could avoid this problem 
(as $\delta \sim \tau^{-1}$); however, 
as one can see from Fig.~\ref{fig_1}, for $\tau > 5.0$
the negative tails of $\zeta_\tau(p)$ are not accessible to our computational
resources. The computation of the finite time corrections 
is mandatory if one aims to test
the fluctuation relation at high values of the external driving force.

\vskip10pt
\noindent
{\it Summary of the data analysis} - To conclude, we summarize the procedure
we follow to analyze the data of a given simulation run:
\begin{enumerate}
\item we determine a value of $\tau$ such that $\zeta_\tau(p)$ appear to
be close to the asymptotic
limit $\zeta_\infty(p)$;
\item we determine the maximum $\widetilde p$ of $\zeta_\tau(p)$ 
by a sixth-order polynomial fit around $p=1$, in an
interval as big as possible compatibly with the request
that the $\c^2$ from the fit is less than $\sim 10$;
\item we shift the histogram of an integer multiple $a$ of the half 
bin size $\delta p/2$ and
compute the function $\chi(a)$ according to Eq.~\ref{chia}. 
We determine the value $a^*$ 
such that the minimum of $\chi(a)$ is
assumed in the interval $[a^*,a^*+1]$: 
the consistency of our analysis requires that
$\delta = 1 - \widetilde p$ and $\d_0=(a^*+0.5)\d p/2$ 
coincide within their errors
(\ie $|\d-\d_0|<\d p$);
\item The value $\chi^* = \min [\chi(a^*),\chi(a^*+1)]$ is an
upper limit for the value of $\chi$ at the minimum.
The number of bins $\min\{M_{a^*},M_{a^*+1}\}$ involved in 
this estimate will be called $M^*$; 
\item we compute the error $\delta X = 2( \overline X(a^*+1) - 
\overline X(a^*))$.
\end{enumerate}
The relevant quantities $\tau$, $\delta$, $\d_0$, $|\d-\d_0|$, $\d p$, 
$M^*$, $\chi^*$ and $\delta X$ 
for model I are
reported in table~\ref{tab:I} for different values of the external force $E$.

\section{Numerical simulation of model I}

\begin{figure}[t]
\includegraphics[width=.50\textwidth,angle=0]{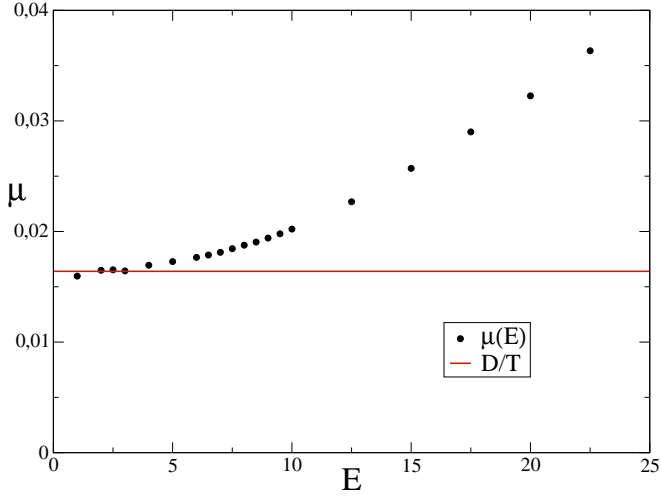}
\caption{Model I: mobility $\mu$ as a function of the driving force $E$. 
The full line is the
equilibrium diffusion coefficient $D$ divided by the temperature. 
Deviations from the
linear response are observed around $E=5$. 
The error bars are of the order of the
dimension of the symbols. Studying $\m(E)$ 
for values of $E$ bigger than those shown in the figure,
one can verify that
the mobility increases
up to a value $\m_{max}$, reached in correspondence of
$E\sim 45$. For values of $E$ bigger than $E\sim 45$, the mobility
begins to decrease essentially following the limiting curve 
$T J_T/(NE)$, where $J_T=\sqrt{T(d-1/N)}N/T$
is the maximum allowed value of the current (saturation value).}
\label{fig_6}
\end{figure}

\begin{table}[b]
\begin{tabular}{r|ccccccccc}
\hline
$E$ & $\tau$  & $\sigma_+$ & $\delta$ & $\d_0$ & $|\d-\d_0|$ & 
$\d p$ & $M^*$ & $\chi^*$ & $\delta X$ \\
\hline
2.5 & 5.0 & 0.194 & 0.272 & 0.183 & 0.089 & 0.244 & 43 & 2.2 & 0.24 \\
5.0 & 5.0 & 0.810 & 0.215 & 0.139 & 0.076 & 0.187 & 20 & 3.5 & 0.22 \\
7.5 & 4.0 & 1.945 & 0.197 & 0.116 & 0.081 & 0.116 & 18 & 2.8 & 0.18 \\
10.0 & 2.5 & 4.044 & 0.262 & 0.151 & 0.111 & 0.122 & 17 & 4.4 & 0.20 \\
12.5 & 2.5 & 7.090 & 0.257 & 0.137 & 0.120 & 0.111 & 8 & 3.5 & 0.28 \\
\hline
\end{tabular}
\caption{Model I: results of the data analysis 
for some selected values of $E$. 
All the quantities are defined in section~\ref{sec:IV}. 
For $E > 12.5$ the negative tails of the distribution 
are not accessible to our numerical simulation.}
\label{tab:I}
\end{table}

We will now discuss systematically the numerical data 
obtained from the simulation
of model I (defined in section~\ref{sec:III}) 
at different values of the driving
force $E$. In Fig.~\ref{fig_6} we report the {\it mobility}
$\mu(E) = T \langle J \rangle_E / (N E)$, {\it i.e.} 
the l.h.s. of Eq.~\ref{GKestesa} times $T/N$, as 
a function of $E$. The current $J(\ul p,\ul q)$ has 
been defined in Eq.~\ref{Jcolor}.
From the Green-Kubo relation, Eq.~\ref{GK},
we have \cite{EM90}
\beq
\lim_{E \rightarrow 0} \mu(E) = \frac{D}{T} \ ,
\eeq
where $D$ is the equilibrium diffusion coefficient,
\beq
D= \lim_{t\rightarrow \infty} \frac{1}{2Nd} \sum_i 
\langle |q_i(t)-q_i(0)|^2 \rangle_{E=0}
\eeq
Deviations from the linear response are observed and 
$\mu(E) \sim D/T + O(E^2)$ above $E=5$.

\begin{figure}[t]
\includegraphics[width=.40\textwidth,angle=0]{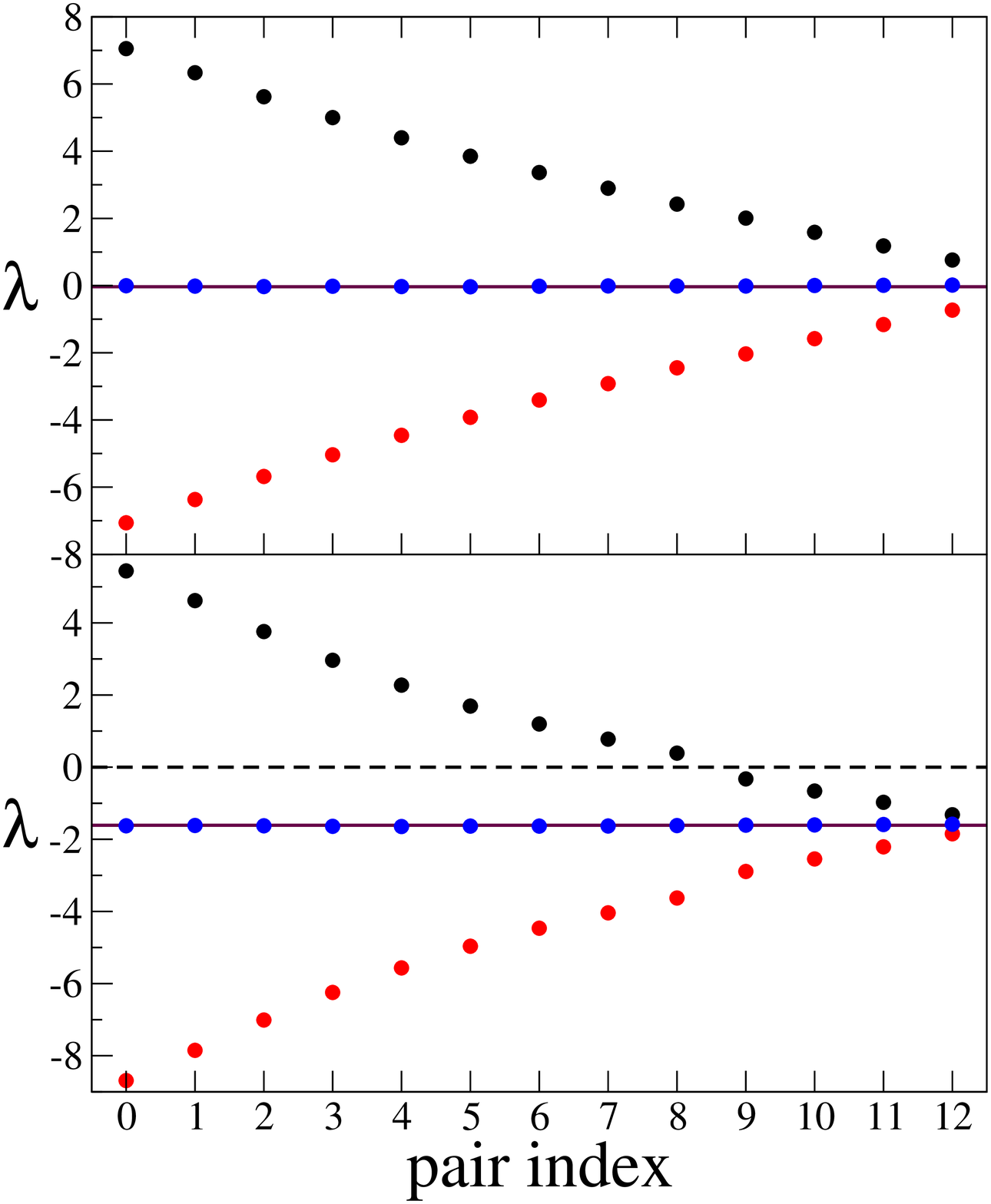}
\caption{Model I: Lyapunov exponents for $E=5$ (top) and for $E=25$ (bottom).
For each panel, the upper and lower dots are the two paired 
exponents $\lambda^{(+)}_j$
and $\lambda^{(-)}_j$, and the
middle dot is their average $(\lambda^{(+)}_j+\lambda^{(-)}_j)/2$.
The full line is $\sigma_+/2Nd$, the dashed line is at $\lambda=0$.
}
\label{fig_7}
\end{figure}

In table~\ref{tab:I} we report the main parameters 
that result from the data analysis
(as discussed in the previous section) for some selected values of $E$.
The value $|\d-\d_0|$ is always less than $\d p$, consistently with our
discussion above, except for $E=12.5$ where, however, the relative difference 
between the two quantities is small ($\sim 9\%$). 
It can be noted that
$\d$ is systematically bigger than $\d_0$. 
This could be due to the fact that  
the error terms $O((p-1)^2/\t)$ or $o(1/\t)$ that we are discarding
likely produce a systematic shift in $\d$ or in $\d_0$; or that
the velocity of convergence of $\z_\t(p)$ is not the same on the negative or 
on the positive side (because numerically is much more difficult to
observe big negative fluctuations of $\s$ than the positive ones -- 
and the Fluctuation Relation provides a quantitative estimate
of the relative probabilities). 
At the moment, because of the level of precision of our
simulations, we are not able to investigate this problem in more 
detail, see also Remark (3) in section~\ref{sec:IID}.
On increasing the value of $E$, we are forced to decrease 
the value of $\tau$ we use for
the analysis as, for longer $\tau$, 
the negative tail of the distribution $\zeta_\tau(p)$
becomes unobservable. This can be seen as the number 
$M^*$ of bins used for the computation of
$\chi$ decrease on increasing $E$; 
above $E=12.5$ it is impossible to find a value of
$\tau$ such that $\zeta_\tau(p)$ 
is close to the asymptotic limit and the negative tail
is observable. Thus, the fluctuation relation 
cannot be tested above $E=12.5$ with
our computational power. However, we are able to 
check the fluctuation relation in
the region $E > 5$ where deviations from the 
linear response are observed. Moreover,
the estimated distributions $\zeta_\io(p)$ are very similar to the one reported
in Fig.~\ref{fig_3}: in particular, they are not Gaussian 
in the investigated
interval of $p$ (also for $E < 5$, in the linear response regime).

Finally, in Fig.~\ref{fig_7} 
we report the measured Lyapunov exponents of the model for $E=5$ 
and $E=25$. For this system, the Lyapunov exponents are 
known to be paired \cite{SEC98,SEI98,DM96}
like in Hamiltonian systems
and the average of each pair is a constant equal to $\sigma_+/2Nd$.
For $E=5$, each pair is composed of a negative and a positive exponent.
This means that the attractive set is dense in phase space \cite{BGG97,Ga99b}
and the chaotic hypothesis is expected to apply to the system yielding a slope
$X=1$ in the fluctuation relation, as confirmed by our numerical data. The same
happens up to $E \sim 20$. 
Above $E=20$, there is a number $D$ of pairs composed by two negative exponents
(for $E=25$ we get $D=4$, see Fig.~\ref{fig_7}).
In this situation, the slope $X$ in the fluctuation relation is expected to be
given by $X = 1 - D/Nd$ \cite{BG97,Ga99b}. 
Thus, for $E=25$ one expects $X \sim 0.75$. Unfortunately,
as discussed above, above $E=12.5$ we did not observe negative fluctuations
of the entropy production, and this prediction could not be tested in
our simulation.

\section{Numerical simulation of model II}

\begin{figure}[t]
\includegraphics[width=.50\textwidth,angle=0]{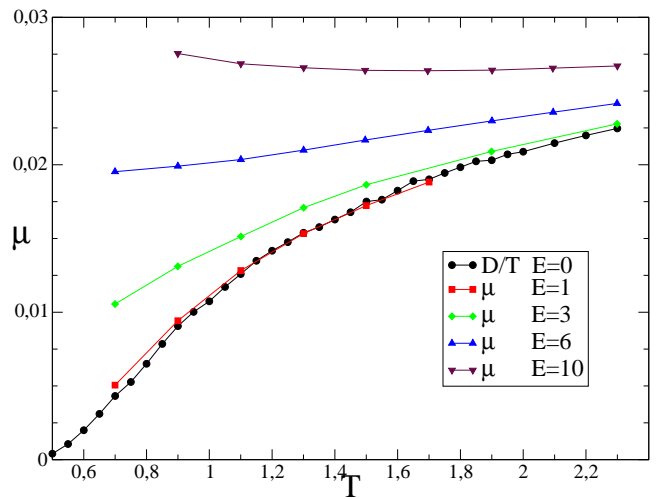}
\caption{Mobility as a function of the temperature $T$ and of the 
driving force $E$ for model II. The circles correspond to the
equilibrium diffusion coefficient divided by the temperature.
Deviations from the linear response are observed for $E \geq 3$;
they become larger on lowering the temperature, 
as $D \rightarrow 0$.
}
\label{fig_8}
\end{figure}

\begin{figure}[t]
\includegraphics[width=.50\textwidth,angle=0]{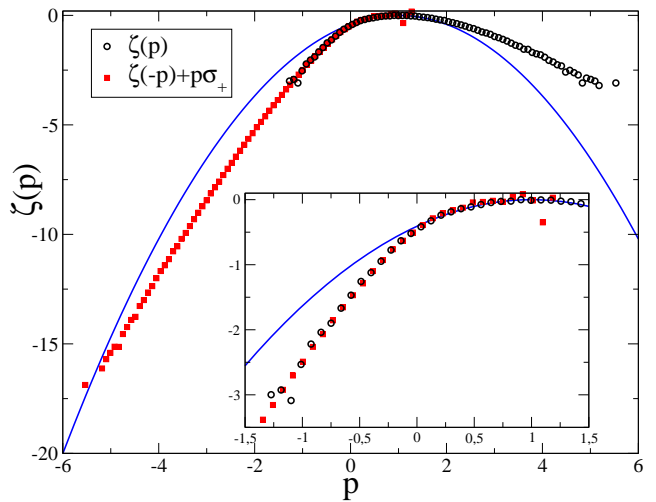}
\caption{The estimate of the function $\zeta_\io(p)$ (open circles) for
model II with $T=1.1$ and $E=3$. In the same
plot $\zeta_\io(-p) + p\sigma_+$ (filled squares) is reported. 
In the inset, the
interval $p\in[-1.5,1.5]$ where the data overlap is magnified. The full
line is the Gaussian approximation, $\zeta_\io(p) = 
\frac12\zeta_\io^{(2)}(p-1)^2$.
The data have been obtained from the histogram of $\pi_\tau(p)$ with $\tau=2.5$
(see table~\ref{tab:II}).
}
\label{fig_9}
\end{figure}

\begin{table}[b]
\begin{tabular}{rr|ccccccccc}
\hline
$T$ & $E$ & $\tau$  & $\sigma_+$ & $\delta$ & $\d_0$ & $|\d-\d_0|$ &
$\d p$ & $M^*$ & $\chi^*$ & $\delta X$ \\
\hline
0.9 & 1 & 3.0 & 0.209 & 0.453 & 0.334 & 0.119 & 0.223 & 68 & 1.9 & 0.19 \\
0.9 & 3 & 3.0 & 2.615 & 0.286 & 0.264 & 0.024 & 0.132 & 15 & 1.0 & 0.23 \\
\hline
1.1 & 1 & 4.0 & 0.233 & 0.231 & 0.126 & 0.105 & 0.126 & 79 & 1.7 & 0.24 \\
1.1 & 3 & 2.5 & 2.493 & 0.217 & 0.238 & 0.021 & 0.087 & 30 & 1.0 & 0.12 \\
1.1 & 6 & 1.5 & 13.32 & 0.113 & 0.230 & 0.117 & 0.092 & 7 & 1.1 & 0.21 \\
\hline
1.5 & 1 & 3.0 & 0.230 & 0.179 & 0.140 & 0.039 & 0.140 & 86 & 0.9 & 0.13 \\
1.5 & 3 & 2.5 & 2.227 & 0.145 & 0.123 & 0.022 & 0.082 & 33 & 4.7 & 0.18 \\
1.5 & 6 & 0.5 & 52.14 & 0.074 & 0.130 & 0.056 & 0.052 & 11 & 0.6 & 0.10 \\
\hline
1.7 & 1 & 3.0 & 0.221 & 0.127 & 0.141 & 0.014 & 0.283 & 49 & 1.0 & 0.26 \\
\hline
1.9 & 3 & 2.5 & 1.981 & 0.106 & 0.122 & 0.016 & 0.122 & 26 & 0.8 & 0.12 \\
1.9 & 6 & 0.4 & 43.52 & 0.078 & 0.126 & 0.048 & 0.085 & 14 & 1.7 & 0.11 \\
1.9 & 10 & 0.2 & 139.0 & 0.079 & 0.135 & 0.056 & 0.039 & 7 & 0.8 & 0.10 \\
\hline
2.1 & 6 & 0.4 & 40.48 & 0.074 & 0.110 & 0.036 & 0.110 & 11 & 1.0 & 0.15 \\
\hline
\end{tabular}
\caption{Model II: results of the data analysis 
for some selected values of $T$ and $E$. 
All the quantities are defined in section~\ref{sec:IV}.
}
\label{tab:II}
\end{table}

Model II differs from model I in the dimension $d=3$, in the larger
number of particles $N=20$, and because it is a binary mixture of
two types of particles.
Binary mixtures are frequently used as models for numerical simulations 
of supercooled liquids as they avoid crystallization also 
at very low temperature on the "physical" time scales (\ie on the time scales
of numerical experiments); for these systems, at low temperature 
deviations from the linear response are observed
also for very low values of the external driving force.

In Fig.~\ref{fig_8} we report the equilibrium diffusion coefficient
$D$ (divided by the temperature $T$) and the mobility (for different
values of $E$) as functions of the temperature.
Even though the number of particles is very small, on lowering the 
temperature the systems approaches the supercooled
state and $D$ becomes very small around $T \sim 0.5$.
Slightly above this temperature, \ie around $T=1$, 
strong deviations from the linear response are
observed for $E \geq 3$, where the entropy production $\sigma_+$ is still
close to $0$. Some values of $\sigma_+$ are reported in table~\ref{tab:II}; 
to compare these values with those obtained for model I one should note
that $\sigma_+$ is an {\it extensive} quantity. Thus, the entropy production
{\it per degree of freedom}, $\sigma_+/2Nd$, is much smaller in model II
than in model I. 

In table~\ref{tab:II} the results of the data analysis outlined in 
section~\ref{sec:IV} are reported. For $E \leq 6$ we obtain a very good 
agreement of the data with the predictions of the fluctuation relation 
and with the  theory of finite time 
corrections discussed in section~\ref{sec:II}.
For $E=10$ it is very difficult to observe negative fluctuations of $p$
with our computational power; see {\it e.g.} the result of the 
analysis for $E=10$ and
$T=1.9$, where only $M^*=7$ bins where available and we were forced to use
$\tau=0.2$, of the order of the mixing time $\tau_0$.
In Fig.~\ref{fig_9} we report the estimated function $\zeta_\io(p)$ obtained
for $T=1.1$ and $E=3$ from the data with $\tau=2.5$. Strong deviations from
the Gaussian behavior are observed in the accessible range of $p$ (see the
inset of Fig.~\ref{fig_9}). A similar behavior
of $\z_\io(p)$ is observed in correspondence of all the values
of $E$ and $T$ we investigated (those listed in Table II):
in particular in all these cases highly non Gaussian behaviors
are observed in the accessible range of $p$. 

The Lyapunov spectrum for this system is very similar to the one reported
in the upper panel of Fig~\ref{fig_7}. Pairs of two negative exponents were
observed only for $E=10$ at $T \leq 1.3$, where, as in the case of model I,
$\sigma_+$ is too large to allow for a 
verification of the modified fluctuation relation expected in this case,
see the discussion at the end of section V.

\section{Conclusions}

We tested the fluctuation relation, in our opinion quite successfully, 
in a numerical simulation of
two models of interacting particles subjected to an external nonconservative
force and to a reversible mechanical thermostat. 
Our data satisfy the fluctuation relation with a $\chi \leq 3$
and an accuracy of the order of $20 \%$ also for very large values of the
driving force, where strong deviations from the linear response are observed, 
and where the large deviation function is strongly non-Gaussian.
The comparison of our numerical data with the predictions of the 
fluctuation relation is done by taking into account the 
(lowest order) finite time
corrections to the distribution function for the fluctuations of
the phase space contraction rate. This is crucial: if we did not 
take into account such corrections the fluctuation relation would
be violated within the precision of our experiment.

In order to compute the finite time corrections, 
we proposed an algorithm which allows to reconstruct the asymptotic 
distribution function from measurable quantities at finite time, within
a given precision. Our theory of the corrections relies on the 
symbolic representation of the chaotic dynamics, therefore it is 
applicable if one accepts the Chaotic Hypothesis.

Our interpretation of the numerical results is that the 
{\it chaotic hypothesis} can be applied to these
systems, also very far from equilibrium, and in particular
the fluctuation relation is 
satisfied even in regions where its predictions measurably differ 
from those of linear response theory.

Our theory of finite time corrections for the analysis of our
numerical data could in principle be of interest for real experimental
settings where non Gaussian fluctuations for the entropy production
rate are observed, see \cite{FM04,CL98}.

However it should be stressed that in a real experiment there are some
technical differences with respect to our numerical simulation which
could in some cases make 
inapplicable 
our analysis, namely:
\\ 
{\it (i)} usually the noise in the large deviation function for the
entropy production rate in a real experiment is much bigger than in a
numerical experiment, and it is likely that the translation in 
Eq.~\ref{etatau} computed as the ratio $\z^{(3)}/(\z^{(2)})^2$ is not
measurable within an error of some percent;
\\
{\it (ii)} usually in a real experiment the accessible time scales
are naturally much bigger than the microscopic ones so that, if the
negative fluctuations of the entropy production rate are observable at all,
one is automatically in the asymptotic regime, where the finite time
corrections should be negligible;
\\
{\it (iii)} a usual problem in a realistic setting is that there is no
clear connection between the ``natural'' thermodynamic entropy
production rate $\dot s=W/T$ ($W$ is the work of the dissipative
external forces and $T$ is the temperature) and the microscopic phase
space contraction rate, for which a slope $X=1$ in the fluctuation
relation $\z(p)-\z(-p)=X\sigma_+p$ is expected; 
so, often one measures an $X\neq 1$ and correspondingly one {\it
defines} an effective temperature $T_{eff}=T/X$ giving a natural
connection between the effective thermodynamic entropy production rate
$\dot s_{eff}=W/T_{eff}$ and the phase space contraction rate, see
\cite{CL98,FM04,Ga04};
in such a situation (where an adjustable parameter $X$ appears) it makes 
no sense to apply our analysis, which is sensible only if one wants to compare
the experimental data with a sharp prediction about the slope $X$ in the
fluctuation relation.

A big open problem we are left with is trying to understand 
how the fluctuation relation is modified for values
of the driving force so high that the attractive set is no longer dense in phase
space. It is expected, \cite{BG97},
that in such a case 
$\z_\io(p)-\z_\io(-p)$ is still linear, but the slope is $X\s_+$, with 
$X$ given by the ratio of the dimension of the attractive set and
of that of the whole phase space. An estimate of such quantity can be given 
via the number of negative pairs of exponents in the Lyapunov spectrum 
\cite{BG97,Ga99b}.
Unfortunately negative pairs begin to appear in the Lyapunov spectrum
only for values of the external force so high that no negative fluctuations
are observable anymore.
We hope that future work will address this point.

\appendix
\section{A Limit Theorem}
\label{app:A}

In this section we prove Eq.~\ref{24}--\ref{23}. 
We reproduce in detail the proof 
in the case $p$ is the average 
of independently distributed discrete variables $\s_i^\e$, assuming
values in $\e\ZZZ$, for some small mesh parameter $\e$; then we discuss
how this can be applied and adapted to the situation
considered in section IIC and subsequent sections.

Let us introduce some definitions. Let $\s_i$, $i\in\NNN$, 
be independent continuous random variables with identical 
distributions $\p(d\s_i)$ with positive variance $\d\s^2>0$, supported
on the finite interval $[s_-,s_+]$. Let us assume that $\p(d\s_i)$
gives positive probability to any finite interval contained in $[s_-,s_+]$.
Let $\p_\l(d\s)$ be the weighted distribution 
$\p_\l(d\s)=e^{-\l\s}\p(d\s)/\int e^{-\l\s}\p(d\s)$ and let us define
$z_\io(\l)=-\log\int e^{-\l\s}\p(d\s)$ and $\s_+=z_\io'(0)$.
Note that the assumption 
that $\p(d\s_i)$ gives positive probability to an interval of $\s$
in $[s_-,s_+]$ implies that for any finite
$\l$ also $\p_\l(d\s)$ has positive variance $-z_\io''(\l)>0$. 

Also, given $\e>0$ (with the property that $s_+-s_-=N_\e\e$ for some integer 
$N_\e$), 
let us consider the discretization of 
$\s_i$ on scale $\e$, call it $\s_i^\e$: $\s_i^\e$ will be a discrete variable 
assuming the values 
$s_k^\e\defin s_-+(k-\frac{1}{2})\e$, $k=1,\ldots,N_{\e}$, 
with probabilities $\p^\e(s_k^\e)=
\Prob(\s_i^\e=s_k^\e)=\int_{s_k^\e\pm\frac{\e}{2}}
\p(d\s)$. The assumption that $\p(d\s_i)$
gives positive probability to any finite interval contained 
in $[s_-,s_+]$ implies that $\p^\e(s_k^\e)>0$ for any $\e$ and $k$.
Let also $z_\e(\l)=-\log\sum_{k=1}^{N_\e}e^{-\l s_k^\e}\p^\e
(s_k^\e)$ and $\p^\e_\l(s_k^\e)=\p^\e(s_k^\e)e^{-\l s_k^\e+z_\e(\l)}$.
Note that, since $\p^\e(s_k^\e)>0$ for any $k$, for any finite $\l$ one has
$-z_\e''(\l)>0$.

If $p_\t^\e=\frac{1}{\t\s_+}\sum_{i=1}^\t\s_i^\e$ and $\P_\t(\e;I)$ is the 
probability 
that $p_\t^\e$ belongs 
to the finite interval $I$, the following theorem holds.\\
\\
{\bf Theorem:} 
{\it Given a finite interval $I\subset (s_-,s_+)$, let $\s_i^\e$, $\p^\e$ and 
$\P_\t(\e;I)$ be defined as above. Then, for a sufficiently small $\e>0$,
there exists an analytic "rate function" $\widetilde\z_\t(p)$ such that
\beq\label{A1.1}
\lim_{\t\to\io}\frac
{\P_\t(\e;I)} 
{\int_{I} dp e^{\t\wt\z_\t(p)}}
=1\;.\eeq
$\widetilde\z_\t(p)$ is defined by:
\beq\label{A1.2}\begin{split}
&\wt\z_\t(p)+\frac{1}{\t}\log\Big[\frac{\sinh[\e\l_p^\e/(2\s_+)]}
{\e\l_p^\e/(2\s_+)}
\Big]=\z_\t^\e(p)\\
&\z_\t^\e(p)=-z_\e(\l_p^\e)+\l_p^\e p\s_+-\frac{1}{2\t}
\log[\frac{2\p}{\t}\Big(-\frac{z_\e''(\l_p^\e)}{\s_+^2}\Big)]\end{split}
\eeq
and $\l_p^\e$ is the inverse of $p(\l)=z_\e'(\l)/\s_+$. 
The function $\z_\t^\e(p)$ has the 
following property: if $\D\subset I$ is an interval of 
size $\frac{\e}{\t\s_+}$ around
a point $p_\D$, then:
\beq\label{A1.1a}
\lim_{\t\to\io}\frac
{\P_\t(\e;\D)}{|\D|e^{\t\z_\t^\e(p_\D)}}=1\eeq
}
\\
{\bf Proof}
Let us introduce the 
auxiliary variable $q={1\over\t\s_+}\sum_{i=1}^\t\h_i$,
where $\h_i$ are i.i.d. discrete random variables, with distribution 
$\p_\l^\e(s_k^\e)$. Let us call $\P_\t^\l(\e;q_0)$ the probability that 
$q$ assumes the value $q_0\in I$, with $q_0\s_+=s_k^\e/\t$ for some $k\in\NNN$,
and note that
$\P_\t^0(\e;q_0)$ is identical to the probability that $p_\t=q_0$.
By definition $\P_\t^\l(q_0)$ and $\P_\t^0(q_0)$ are related by:
\beq\label{A1.5}
\P_\t^\l(\e;q_0)={e^{-\l q_0\s_+\t}\P_\t^0(\e;q_0)\over 
\big[\sum_k e^{-\l s_k^\e} \p^\e(s_k^\e)\big]^\t}
\eeq
Now, a local form of central limit theorem 
(Gnedenko's theorem, see pag. 211 of \cite{Fi63}) tells us that, 
if $q$ is localized near its mean value, that is if 
$|q\s_+-z_\e'(\l)|\le \frac{M\e}{\t}$
for some finite $M$, then $\P_\t^\l(\e;q_0)$ is asymptotically equivalent
to the Gaussian with mean $z_\e'(\l)$ and variance 
$-z_\e''(\l)$, in the sense that
\beq\label{A1.6}\P_\t^\l(q_0)=\frac{\e}{\sqrt{2\p\t(-z_\e''(\l))}}e^{-
\frac{(q_0\s_+-z'_\e(\l))^2}{2(-z_\e''(\l))}\t}(1+o(1))\;,\eeq
for any $q_0$ s.t. $|q\s_+-z_\e'(\l)|\le \frac{M\e}{\t}$
\footnote{\label{central} Note that Gnedenko's Theorem is {\it different}
from the usual central limit theorem, stating instead that for 
$|q\s_+-z_\e'(\l)|\le \frac{C}{\sqrt\t}$ ($C$ big) the sums of 
$\P_\t^\l(\e;q)$ over intervals of amplitude $\frac{1}{\sqrt\t}$ contained in 
$|q\s_+-z_\e'(\l)|\le \frac{C}{\sqrt\t}$
are asymptotically equal to the integrals of the Gaussian 
over the same intervals. 
That is, usual central limit theorem gives informations on the distribution
in a bigger interval around the maximum, but on a rougher scale.}.

So, given $\l_{q_0}^\e$ s.t. $z_\e'(\l_{q_0}^\e)=q_0\s_+$ (such $\l_{q_0}^\e$ 
exists, is unique
and is an analytic function of $q_0$, by the remark that $-z_\e''(\l)>0$ for
any finite $\l$ and
$z_\e(\l)$ is an analytic function of $\l$), using
Eq.~\ref{A1.6} we see that 
Eq.~\ref{A1.5} can be restated as:
\beq\label{A1.7}\P_\t^0(\e;q_0)=\frac{\e}{\sqrt{2\p\t(-z_\e''(\l_{q_0}^\e))}}
e^{\l_{q_0}^\e q_0\s_+\t-z_\e(\l_{q_0}^\e)}(1+o(1))\eeq
Now, by the definition of $\z_\t^\e(p)$ in Eq.~\ref{A1.2}, we see that 
the r.h.s. of the last equation is equal to 
$\frac{\e}{\t\s_+}e^{\t\z_\t^\e(q_0)}(1+o(1))$.
Finally, the statement of the Theorem follows by the remark that 
\beq\label{A1.8}
\frac{\e}{\t\s_+}e^{\t\z_\t^\e(p_0)}=\int_{p_0-\frac{\e}{2\t\s_+}}^{p_0+\frac{\e}
{2\t\s_+}}dp e^{\t\tilde\z_\t(p)}\Big(1+o(1)\Big)\;.\eeq	
In fact the integral in the r.h.s. of the last equation is given by
\beq \label{A1.9}\begin{split}
&e^{\t\tilde\z_\t(p_0)} \int_{p_0-\frac{\e}{2\t\s_+}}^{p_0+\frac{\e}
{2\t\s_+}}dp e^{\t\tilde\z_\t'(p_0)(p-p_0)}\Big(1+O(\frac{\wt\z_\t''(p_0)\e^2}{\t})
\Big)=\\
&=e^{\t\tilde\z_\t(p_0)} \frac{2\sinh[\tilde\z_\t'(p_0)\e/(2\s_+)]}
{\t\tilde\z_\t'(p_0)}
\Big(1+O(\frac{\wt\z_\t''(p_0)\e^2}{\t})\Big)\end{split}\eeq
and in the last expression one has to note that $\wt\z_\t'(p_0)=[\z_\t^\e]'(p_0)+
O(\frac{1}{\t})=\l_{p_0}^\e+O(\frac{1}{\t}).\ $ \qed
\\
\\
A first Remark to be done about the Theorem above is that, in order
to define a ``universal'' rate function in terms of quantities 
depending only on 
$z_\io(\l)$ (instead of quantities depending 
on the ``non universal'' function $z_\e(\l)$, 
which explicitly depends on the discretization step $\e$), it would
desirable to perform (in a sense to be precised) 
the continuum limit $\e\to 0$. To this regard, we can note that 
the only point where in the proof above we really used the fact that 
$\e$ is a constant (\ie is independent of $\t$) was in using Gnedenko's 
Theorem, see \cite{Fi63}. However, by a critical analysis of the proof 
of Gnedenko's Theorem, one can realize that it is even possible to 
let $\e=\e_\t$ go to $0$ with $\t$; 
the velocity with which $\e_\t$ is allowed to go to $0$ depends on the 
details of the distribution $\p(d\s)$. 
So we can even study the probability distribution of $p_\t$ on a scale
$\sim \e_\t/\t$: if we introduce bins $\D_\t$ of size $O(\e_\t/\t)$
and we define $\P_\t(\D_\t)$ to be the probability that $p_\t={1\over \t\s_+}
\sum_i\s_i$ belongs to the bin $\D_\t$ centered in $p_0$, we can 
repeat the proof above to conclude that
\beq\label{A1.9a}
\lim_{\t\to\io}\frac
{\P_\t(\D_\t)}{|\D_\t|e^{\t\z_\t(p_0)}}=1\eeq
where $\z_\t$ satisfies the equation:
\beq\z_\t(p)=-z_\io(\l_p)+\l_p p\s_+-\frac{1}{2\t}
\log[\frac{2\p}{\t}\Big(-\frac{z_\io''(\l_p)}{\s_+^2}\Big)]\eeq
and $\l_p$ is the inverse of $p(\l)=z_\io'(\l)/\s_+$. \\
 
Another point to be discussed is that in the Theorem above we assumed 
the $\s_i$ to be independent. 
This is not the case for the variables $\s(S^i\cdot)$ of sec.~\ref{sec:II}. 
However, if, as discussed in Remark (3) of sec.~\ref{sec:IID},
we choose the time unit to be of the order of the mixing time, the variables 
$\s(S^i\cdot)$ have (by construction) a decorrelation time equal to $1$, and the 
analysis of previous theorem can be repeated step by step
in order to construct the probability distribution of 
$p=\frac{1}{\t\s_+}\sum_{i}\s(S^i\cdot)$. The only differences
are that: (1) $\t z_\io(\l)$ should be replaced by $\t z_\t(\l)
=-\log\int 
e^{-\l p\s_+\t}\P_\t(dp)$ throughout the discussion; (2) instead of Gnedenko's
theorem one has to apply a generalization of Gnedenko's to short ranged
Gibbs processes, to be proven via standard cluster expansion
techniques (see for instance \cite{Ga72} for a proof of a generalization
of Gnedenko's theorem to a short ranged Gibbs process
in the context of non critical fluctuations 
of the phase separation line in the 2D Ising model). \\

The conclusion is that, if the bins $\D$ 
in sec.~\ref{sec:IIC} are chosen of size $\e_\t/\t$, the probability of the bin $\D$
centered in $p_\D$ is asymptotically given by $\p(p\in\D)\simeq e^{\t\z_\t(p_\D)}$
(in the sense of Eq.~\ref{24}) and  $\z_\t(p_\D)$ can be interpolated by
an analytic function of $p$ that in fact satisfies Eq.~\ref{23}.

\acknowledgments

We would like to thank Prof.~Joel~L.~Lebowitz for his interest
on this work and for a valuable discussion. Some of this work was
completed while two of us (A.G. and G.G.) were visiting Rutgers University 
under invitation of Prof.~Lebowitz. A.G. was partially supported by 
the NSF grant 4-23421 DMR 01-279-26.

We thank F.~Bonetto for a useful discussion and his suggestions about
the implementation of the numerical algorithm.

F.Z. wish to thank G.~Ruocco and L.~Angelani for many useful 
discussions.

The computations have been performed on the FDT cluster of the INFM-SOFT center
in Rome. We thank S.~Erriu for technical assistance in operating the cluster.
The sources of the program we used for the simulations
can be downloaded from {\it http://glass.phys.uniroma1.it/zamponi}.


\end{document}